\documentclass[twocolumn,tighten]{aastex62}
\usepackage{lineno}
\usepackage{bm}
\usepackage{xspace}
\usepackage{enumitem}
\usepackage{graphicx}
\usepackage{color}
\usepackage{comment}
\usepackage{amsmath}
\usepackage[T1]{fontenc}
\usepackage{soul}
\providecommand{\dodoi}[1]{doi:~\href{http://doi.org/#1}{\nolinkurl{#1}}}

\providecommand{\doarXiv}[1]{arXiv:~\href{https://arxiv.org/abs/#1}{\nolinkurl{#1}}}

\graphicspath{{./}{figures/}}

\usepackage{color}

\graphicspath{{./}{figures/}}
\shorttitle{A Library of Synthetic TDE X-ray Spectra}
\shortauthors{Wen et al.}
\begin{document}
\title{A Library of Synthetic X-ray Spectra for Fitting Tidal Disruption Events}
\correspondingauthor{Sixiang Wen}
\email{S.Wen@astro.ru.nl}
\author[0000-0002-0934-2686]{Sixiang Wen}
\affiliation{Department of Astrophysics/IMAPP, Radboud University, P.O.~Box 9010, 6500 GL, Nijmegen, The Netherlands}
\affiliation{University of Arizona, 933 N. Cherry Ave., Tucson, AZ  85721}

\author[0000-0001-5679-0695]{Peter G.~Jonker}
\affiliation{Department of Astrophysics/IMAPP, Radboud University, P.O.~Box 9010, 6500 GL, Nijmegen, The Netherlands}
\affiliation{SRON, Netherlands Institute for Space Research, Niels Bohrweg 4,
2333 CA, Leiden, The Netherlands}

\author[0000-0002-4337-9458]{Nicholas C. Stone}
\affiliation{Racah Institute of Physics, The Hebrew University, Jerusalem, 91904, Israel}

\author[0000-0001-6047-8469]{Ann I. Zabludoff}
\affiliation{University of Arizona, 933 N. Cherry Ave., Tucson, AZ  85721}

\author[0000-0002-0588-6555]{Zheng Cao}
\affiliation{Department of Astrophysics/IMAPP, Radboud University, P.O.~Box 9010, 6500 GL, Nijmegen, The Netherlands}
\affiliation{SRON, Netherlands Institute for Space Research, Niels Bohrweg 4,
2333 CA, Leiden, The Netherlands}

\begin{abstract}
We present a tabulated version of our slim disk model for fitting tidal disruption events (TDEs). We create a synthetic X-ray spectral library by ray-tracing stationary general relativistic slim disks and including gravitational redshift, Doppler, and lensing effects self-consistently. We introduce the library to reduce computational expense and increase access for fitting future events. Fitting requires interpolation between the library spectra; the interpolation error in the synthetic flux is generally $<10\%$ (it can rise to $40\%$ when the disk is nearly edge-on). We fit the X-ray spectra of the TDEs ASASSN-14li and ASASSN-15oi, successfully reproducing our earlier constraints on black hole mass $M_\bullet$ and spin $a_\bullet$ from full on-the-fly ray-tracing. We use the library to fit mock observational data to explore the degeneracies among parameters, finding that 1) hotter thermal disk and edge-on inclination angle spectra offer tighter constraints on $M_\bullet$ and $a_\bullet$; 2) the constraining power of spectra on $M_\bullet$ and $a_\bullet$ increases as a power-law with the number of X-ray counts, and the index of the power law is higher for hotter thermal disk spectra; 3) multi-epoch X-ray spectra partially break the degeneracy between $M_\bullet$ and $a_\bullet$; 4) the time-dependent level of X-ray absorption can be constrained from spectral fitting. The tabulated model and slim disk model are {\href{https://doi.org/10.25739/hfhz-xn60}{here.}}


\end{abstract}


\section{Introduction}
\label{int}

Tidal disruption events (TDEs) are a once-rare class of transient now discovered at a rapidly increasing pace.  Current optical and soft X-ray wide-field surveys, such as ASAS-SN \citep{Holoien+14, Holoien+16a}, ZTF \citep{vanVelzen+21}, and {\it eROSITA} \citep{Sazonov+21}, are discovering tens of new TDEs per year \citep[see][for recent reviews]{Saxton+20, vanVelzen+20a}, and almost one hundred candidate TDEs have been discovered to date\footnote{At the time of writing, 98 candidate TDE flares are listed at the \href{https://www.tde.space}{Open TDE Catalog}.}.  Near-future surveys in the optical (VRO; \citealt{Ivezic+19}) and ultraviolet ({\it ULTRASAT}; \citealt{Sagiv+14}) will likely find hundreds to thousands more each year \citep{BricmanGomboc20}.

Such a large sample of TDEs raises the prospect of studying supermassive black hole (SMBH) mass and spin demographics.  The canonical TDE forms when a star passes too close to a SMBH in a quiescent galactic nucleus, is torn apart, and then produces a luminous, multiwavelength flare \citep{Hills75, LidskiiOzernoi79, Rees88, EvansKochanek89}.The physical evolution and thermal emission from TDEs is governed by just a small number of free parameters (see \citealt{Rossi+21} for a recent review). For the standard case of a main sequence star disrupted on a nearly-parabolic orbit\footnote{If the disrupted star is not on the main sequence or comes from a non-parabolic orbit, then additional parameters enter the problem, but these types of TDEs are less common \citep{MacLeod+12, Hayasaki+18} and are expected to have very different light curves \citep{MacLeod+13, Hayasaki+13, Hayasaki+18}.}, the five controlling parameters are SMBH mass $M_\bullet$ and stellar mass \citep{Rees88, GuillochonRamirezRuiz13, Ryu+20a, Ryu+20b}, SMBH spin $a_\bullet$ \citep{Dai+15, Hayasaki+16, Wen+20}, orbital pericenter \citep{Stone+13, GuillochonRamirezRuiz13, Dai+15}, and spin-orbit misalignment angle \citep{GuillochonRamirezRuiz15, Hayasaki+16, Liptai+19}.  Other properties of the star, such as its metallicity, rotation rate, and age along the main sequence, can affect the debris evolution, but are unlikely to do so at leading order \citep{GallegosGarcia+18, Kagaya+19, Golightly+19, LawSmith+19, LawSmith+20}. The small number of free parameters associated with the flare suggests that $M_\bullet$ and $a_\bullet$ are usefully encoded in the resulting electromagnetic emission.  

Measuring the distribution of $M_\bullet$ is a path toward quantifying the uncertain low mass tail of the SMBH mass function, including elusive intermediate-mass black holes (IMBHs), and constraining SMBH seed formation scenarios \citep{Volonteri10}. Constraining the SMBH spin distribution would recover the growth history of SMBHs \citep{Berti+08}. The $a_\bullet$ distribution is poorly understood at present, with most existing measurements arising from a single technique: iron K-$\alpha$ reflection spectroscopy \citep{Reynolds14}. The SMBH spin is imprinted on TDE observables in various ways, e.g., relativistic precessions \citep{StoneLoeb12, GuillochonRamirezRuiz15, Hayasaki+16, Liptai+19, Curd21, Andalman+22}, the disk thermal X-ray spectrum \citep{McClintock+06,Done2012}, the TDE rate \citep{Kesden2012}, and quasi-periodic oscillations in the X-ray spectrum \citep{Dheeraj2019}, raising the possibility of obtaining independent SMBH spin constraints, with different uncertainties and biases than those of the status quo.

Until recently, most attempts to measure $M_\bullet$ from TDE observations have relied on optical light curves \citep{Mockler+19, Ryu+20c}, the time evolution of which resembles theoretical mass fallback rates \citep{GuillochonRamirezRuiz13}.  Unfortunately, first-principles models for TDE optical emission do not yet exist. More idealized models have not even converged on the underlying power source; at the time of writing, there is a vigorous debate about whether reprocessed accretion luminosity \citep{Guillochon+14, MetzgerStone16, Roth+16, BonnerotLu20} or shock dissipation \citep{Piran+15, Shiokawa+15} is the dominant power source for the optical photosphere.

The quasi-thermal soft X-ray emission observed in a fraction of TDEs is arguably better understood, as the only existing model for it invokes an origin in the innermost regions of the newly formed accretion flow \citep{Ulmer99, LodatoRossi11}.  While uncertainties about the geometry of this inner accretion flow exist, particularly at early times, the X-ray emitting material is generally matter that has dissipated orders of magnitude of energy. It is therefore reasonable to expect significant circularization.  For example, the recent first-principles simulations of \citet{Andalman+22} find that even over the first six days of mass fallback alone, equatorial material accreting onto the SMBH passes through the innermost stable circular orbit with eccentricities $e \approx 0.4 - 0.8$ (much smaller than typical post-disruption $e \approx 0.99$, and likely to decrease further as the inner disk builds up and circularization progresses).  

In recent work, we have combined Kerr metric ray-tracing \citep{JP11} with general relativistic slim disk models \citep{Sadowski09} to produce synthetic X-ray spectra for the inner regions of TDE disks.  By fitting these models to X-ray observations, we have been able to constrain $M_\bullet$ and $a_\bullet$ for both SMBHs \citep[][hereafter W20]{Wen+20} and IMBHs \citep[][hereafter W21]{Wen+21}.  The IMBH case is particularly interesting as a novel probe of SMBH seed formation scenarios and unexplored mass ranges of ultralight dark matter candidates (W21).  However, a practical weakness of our past work is the need to generate a large grid of computationally expensive synthetic X-ray spectra on the fly, for purposes of parameter estimation.  The computational cost of generating custom spectral databases ($\sim 10^4$ CPU hours for each TDE) would become prohibitive if applied at scale to much greater numbers of X-ray bright TDEs, such as those expected in the {\it eROSITA} era \citep{Khabibullin+14, Jonker+20, Sazonov+21}.

In this paper, we address this problem, and thoroughly explore the parameter space of relativistic slim disk models, by producing a large tabulated library of synthetic X-ray spectra. These 
spectra are derived from our original full ray-tracing model (W21). In \S \ref{Methodology}, we review the underlying theoretical model and key assumptions.
In \S \ref{library}, we present the tabulated library of synthetic X-ray spectra. We fit two previously studied TDEs 
by interpolating over this library
in \S \ref{asassns}.  In \S \ref{mock_fitting}, we explore the different types of parameter degeneracy that arise in the multi-dimensional parameter space of X-ray bright TDE disks, with the overall goal of determining which types of observations can constrain different underlying physical parameters.  Finally, we conclude in \S \ref{conclusions}.  So that this library is widely usable as the TDE field transitions to large population studies, we have made it publicly accessible \footnote{\url{https://doi.org/10.25739/hfhz-xn60}}.

\section{Methodology}
\label{Methodology}
Our model is based on the work of W20 and W21, who used a general relativistic stationary slim disk model \citep{Abramowicz1996,Sadowski09}  employing the ray-tracing code from \citet{JP11} to calculate the synthetic spectra as seen by an observer. Here, we review the high-level details of the local X-ray emission and ray-tracing procedure. For the full details on relativistic slim disk solutions and the ray-tracing techniques we use, we refer the reader to W20, W21, and the aforementioned citations in this section. 

In order to calculate the local X-ray spectrum of the disk from the disk flux $F(r)$, we follow \citet{ST95}, assuming a color-modified blackbody is emitted at each annulus of the disk. The specific intensity of the radiation, with frequency $\nu$, can be expressed as, 
\begin{equation}
\label{Iv}
 I_{\rm d}(\nu)=\frac{2h\nu^3c^{-2}f_{\rm c}^{-4}}{\exp(h\nu/k_{\rm B} f_{\rm c}T_{\rm e})-1}.
\end{equation}
Here $h$ is the Planck constant and $k_{\rm B}$ is the Boltzmann constant; $f_{\rm c}$ is commonly called the spectral hardening factor. $T_{\rm e}$ is the local effective temperature of an annular ring of the disk (assumed to be axisymmetric), which is determined by $T_{\rm e}=(F(r)/2\sigma)^{1/4}$ (where $F(r)$ is the flux from the disk at radius $r$ and $\sigma$ is the Stefan-Boltzmann constant).
We use $f_{\rm c}$ values as estimated by the radiative transfer calculations of \citet{DE19}, and which are parametrized as:
\begin{eqnarray}
\nonumber 
f_{\rm c}=1.74+&&1.06(\log_{10} T_{\rm e}-7)-0.14[\log_{10} Q(r)-7] \\ 
&&-0.07\{\log_{10}[\Sigma(r)/2]-5 \}.
\label{fc2}
\end{eqnarray}
Here $\Sigma(r)$ and $Q(r)$ are the surface density and the strength of the vertical gravity at each annulus of the disk, respectively. Although this parametrization is fitted to the best theoretical models for $f_{\rm c}$ in the present literature, we note that they have only been tested/calibrated for accretion rate between 0.01 to 1 times the Eddington accretion rate, and only for non-spinning BHs. Throughout this paper, the Eddington accretion rate is given by
\begin{equation}
    \dot M_{\rm Edd}=1.37 \times 10^{21} ~{\rm kg~s}^{-1} \eta^{-1}_{-1} M_6,
\end{equation}
where $\eta$ is the radiative efficiency of accretion, $\eta_{-1}=\eta/0.1$, $M_6$ is the SMBH mass in units of a million solar masses $M_6=M_\bullet/(10^6M_\odot)$.
Hereafter, we refer to accretion rate in dimensionless Eddington units, i.e., $\dot m= \dot M/ \dot M_{\rm Edd}$, and we assume $\eta=0.1$ in all future references to Eddington ratios $\dot{m}$ (even though the {\it actual} radiative efficiency of our slim disk models can be significantly different).

We trace light rays as null geodesics from the image plane to the surface of the disk, which is truncated at an outer radius $R_{\rm out}$ and an inner radius $R_{\rm in}$. We set $R_{\rm out}$ to two times the tidal radius $R_{\rm t}$ \citep{Hills75}, 
\begin{align}
    R_{\rm t} = R_\star \left( \frac{M_\bullet}{M_\star} \right)^{1/3} \approx ~47 R_{\rm g} M_6^{-2/3}\left(\frac{M_\star}{M_\odot} \right)^{-1/3} \left(\frac{R_\star}{R_\odot} \right).
\end{align}
Here, $R_{\rm g} = GM_\bullet / c^2$ is the gravitational radius and $M_\star$ and $R_\star$ are the mass and radius of the disrupted star, respectively. In calculating $R_{\rm t}$ (to compute $R_{\rm out}$), we assume that the disrupted star has a solar mass and radius. When $2R_{\rm t} > 600 ~R_{\rm g}$, we reset it to be $R_{\rm out}=600 R_{\rm g}$. The error on the luminosity caused by this choice of $R_{\rm out}$ is always $<1\%$ (W21), because the outer disk is too cold to produce a significant X-ray luminosity\footnote{Note that this situation ($R_{\rm out} \gtrsim 600 R_{\rm g}$) only arises for relatively small IMBHs.}. On the other hand, when $M_\bullet> 4.2 \times 10^7 M_\odot$, then $R_{\rm out}$ will usually be smaller than 15 $R_{\rm g}$. For such a case, the choice of $R_{\rm out}$ can affect the spectrum significantly. As the SMBH mass $M_\bullet$ increases, $R_{\rm out}/R_{\rm g}$ will decrease quickly. If $R_{\rm out}<15 R_{\rm g}$, it is reset to $R_{\rm out}=15 R_{\rm g}$.
Due to a singularity in the calculation of $f_{\rm c}$ for disk radii smaller than  the innermost stable circular orbit (ISCO), we set $R_{\rm in}$ to the value of the ISCO. 
The error on the luminosity caused by different choices of inner edge is small\footnote{Regions inside the ISCO contribute little to the total X-ray flux due to (i) strong gravitational redshift; (ii) small emitting area; and (iii) weak spectral hardening.  See also \citet{Zhu+12} for simulations addressing this point in the context of low-to-moderate Eddington ratios.}. A test in W21 shows that the error is $< 2\%$ for different choices of spin and accretion rate for an intermediate mass BH with inclination of $45^\circ$. W21 also finds that, for an accretion disk around a fast spinning BH,  regardless of BH mass, the error is less than $0.5\%$.

If the peak accretion rate for TDEs is highly super-Eddington, the disk will be geometrically thick ($H/R\sim 0.5$, where $H$ is the half disk thickness and $R$ is the disk radius). When the disk inclination is close to edge-on, the edge of the disk will block most X-rays from the inner disk (W20).
When the photon's trajectory intersects the disk at any point during ray tracing, we stop further tracing and assume the photon is emitted from the first intersection point. In addition, we assume that the 
outer rim
of the disk behaves as a single temperature color-corrected blackbody, with an effective temperature equal to that of the standard annular effective temperature of the outermost annulus.

To summarize, in calculating the synthetic X-ray spectrum, we have assumed that:
\begin{enumerate}

\item The inner disk is circular and equatorial.

\item The disk structure is determined assuming a zero-torque inner boundary condition, large optical depth, and no self-irradiation.  It is computed using the general relativistic slim disk model of \citet{Sadowski09}, which is similar to the classic Shakura-Sunyaev and Novikov-Thorne thin disk models, but differs in its inclusion of (i) advective cooling and (ii) sub-Keplerian angular momentum when $H/R$ is large \citep{Abramowicz+88}.  These features allow slim disks to cover the super-Eddington accretion rates that may be present during the early stages of some TDEs.
  
\item The local X-ray emissivity of the disk is consistent with modified multicolor blackbody radiation, e.g., Eq.~\eqref{Iv}. The spectral hardening factor is calculated using Eq.~\eqref{fc2}.
  
\item The X-rays are generated at the finite-height photospheric surface of the disk, which we assume to be equal to its scale height $H$. If a photon's trajectory intersects the disk after emission, we assume that it will be absorbed completely.  Photons follow null geodesics that can be strongly lensed by the Kerr metric they propagate through.
  
\item The inner edge of the disk is set to be the innermost stable circular orbit (ISCO). The outer edge of the disk is twice the tidal disruption radius ($2R_t$). If the outer edge is $>600 R_g$ ($<15R_g$), we reset it to be $600R_g$ ($15R_g$).

\end{enumerate}

\section{Model Spectral Library}
\label{library}
The tabulated synthetic spectra in the library are produced by the ray-tracing model of W21, which includes gravitational redshift, Doppler, and lensing effects self-consistently, as well as the additional effect of angular momentum loss by radiation (which was not considered in W20). In the context of relativistic slim disks, the disk X-ray spectrum can be affected by the following parameters: the black hole mass $M_\bullet$ and spin $a_\bullet$, the disk accretion rate $\dot m$ and inclination $\theta$, viscosity $\alpha$, and spectral hardening factor $f_{\rm c}$. In W20, we showed that the impact of $\alpha$ on the spectrum is small compared with that of $M_\bullet$, $\dot m$, and $\theta$; as a result, 
for the model spectral library, we fix $\alpha=0.1$. 
We calculate $f_c$ following \citet{DE19}; this prescription is calibrated for a range of sub-Eddington accretion disk properties, and extending it to highly super-Eddington accretion rates introduces a systematic uncertainty that should be quantified with future modeling. In general, for the tabulated model, there are only four free disk parameters, $M_\bullet$, $a_\bullet$, $\dot m$, and $\theta$, while $\alpha$ is fixed at $\alpha=0.1$ and $f_{\rm c}$ is estimated by Eq.~\ref{fc2}.  

We introduce two model functions, {\sc slimd} and {\sc slimdz}, for spectral fitting. {\sc slimd} fits the spectra for a given luminosity distance and redshift, while {\sc slimdz} fits the spectra by calculating the luminosity distance from the redshift with a flat cosmological model. We note that redshift can be a free parameter for {\sc slimdz}. For more details about these two model functions, we refer the reader to the menu given with the
synthetic spectral library.

\begin{deluxetable}{cccc}
\tablecaption{Parameter Ranges for Model Grid}
\tablewidth{0pt}
\tablehead{
 $M_\bullet$ [$M_\odot$] & $a_\bullet$  & $\dot m$ [Edd] & $\theta$ [$^\circ$] 
}
\startdata
$[10^3, 9\times 10^7]$ & $[-0.998, 0.998]$ & $[0.05, 600]$ & $[2, 90]$ 
\enddata
\label{tab:prior}
\tablecomments{For calculating the Eddington accretion rate $\dot{M}_{\rm Edd}$, we assume a constant radiation efficiency of $\eta=0.1$ irrespective of the real, $a_\bullet$-dependent efficiency.}
\end{deluxetable}

The library contains model spectra calculated on a grid of parameters $M_\bullet$, $a_\bullet$, $\dot m$, and $\theta$. The size of the grid is $85 \times 33 \times 100 \times 119$. One can find the specific grid values in the model files, and the ranges of values are listed in Table~\ref{tab:prior}. We use linear interpolation to estimate the spectra for parameter values between the grid points. For the parameter $\theta$, we use linear interpolation to estimate the actual flux at each energy bin, while, for the parameters $M_\bullet$, $a_\bullet$, and $\dot m$, we use linear interpolation to estimate the logarithm of the flux\footnote{We make these choices because the flux is roughly $\propto \sin{\theta}$, but also $\propto \exp{(T)}$, and the disk temperature $T$ is function of other disk parameters.}. 
We then compare the flux predicted by interpolating off of grid points on the table with direct, on-the-fly ray-tracing calculations.
The grid is sufficiently fine that the error caused by interpolation is less than $10\%$ for nearly all cases (more on this below). For each grid point, we calculate the spectral luminosity density in 500 photon energy bins spread over the $0.2-10$~keV energy range.

\begin{figure*}[ht!]
\gridline{
\fig{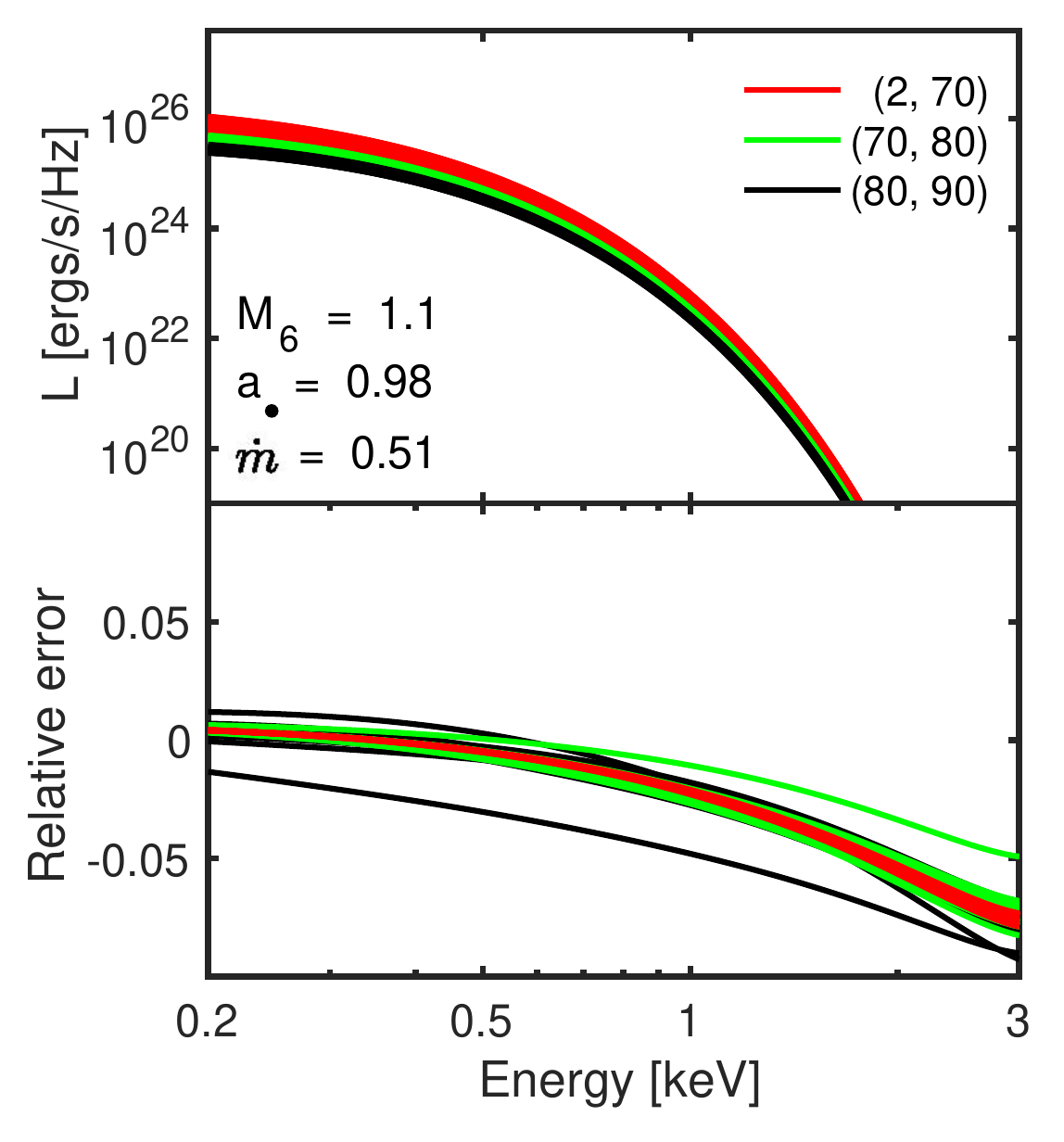}{0.475\textwidth}{}
\fig{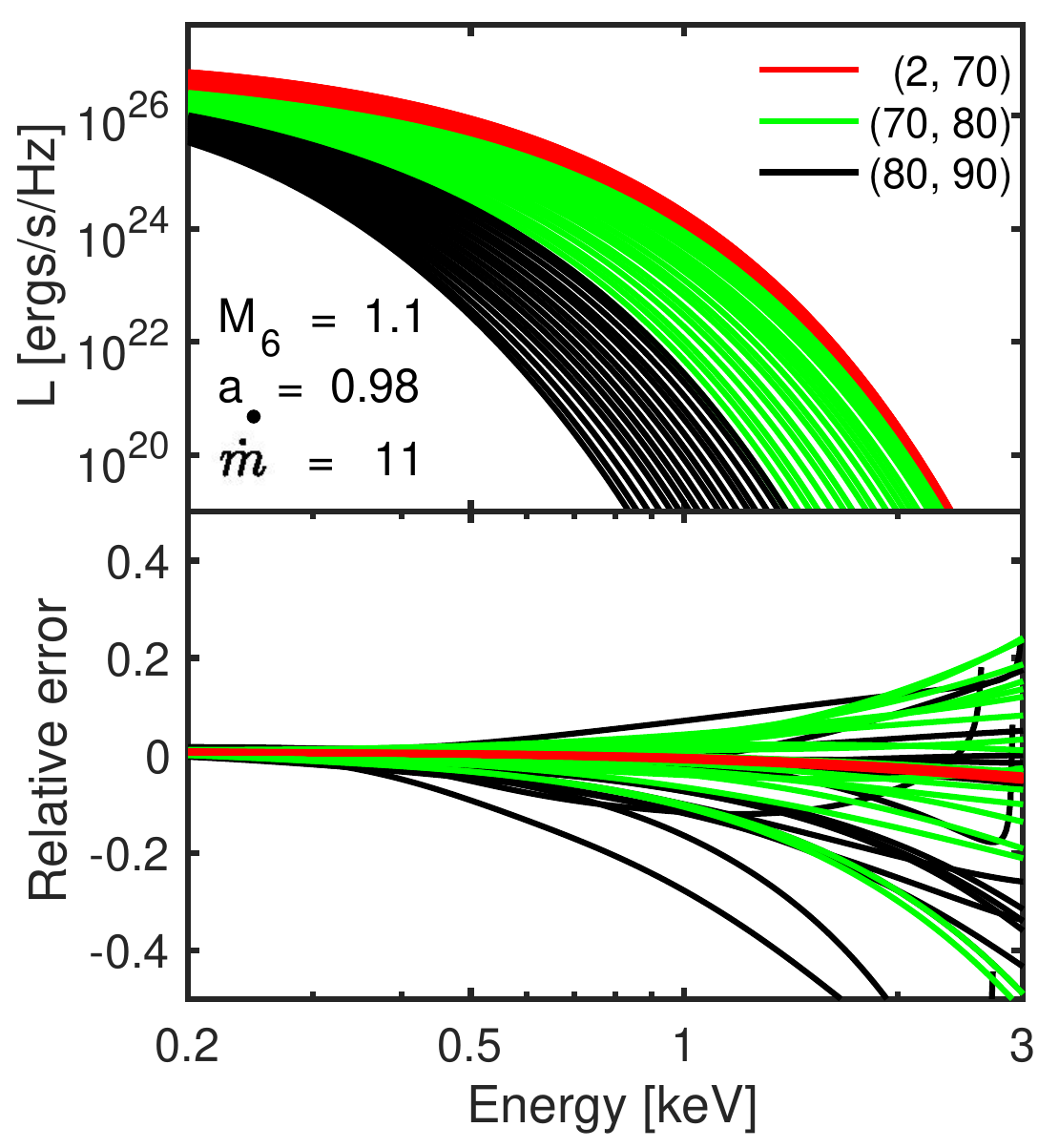}{0.465\textwidth}{}
}
\caption{Synthetic spectra generated through ray-tracing for different inclinations between the disk and our line of sight (upper panels). The solid red lines denote the spectra with inclinations (in the middle of each grid) varying from $2^\circ$ to $70^\circ$, while the green and black solid lines represent inclinations varying from $70^\circ$ to $80^\circ$ and $80^\circ$ to $90^\circ$, respectively. The lower panels show the error on the luminosity density as a function of energy caused by the interpolation between the accurate spectra calculated for the fixed grid parameter values. This error is determined by comparing the spectra calculated using the full ray-tracing calculations ($L_r$) with those determined through interpolation ($L_i$). We define the relative error as: $1-L_i/L_r$. The error increases as the spectra become harder, because the luminosity density decreases exponentially with energy. This figure shows that the error in the luminosity density caused by interpolation is generally $<10\%$ 
over these energies,
while, for high accretion, edge-on cases, the error can be up to $40\%$. }
\label{fig:er1}
\end{figure*}
 
Figure \ref{fig:er1} shows the spectra and the error on the luminosity density caused by the linear interpolation between the grid points. 
For the upper panels, the spectra are calculated with the full ray-tracing code even if their parameters do not lie on the grid points of the library. We also produce spectra for disks with the same parameters, but by using linear interpolation on the library of pre-tabulated spectral models. We compare the full ray-traced spectra with those obtained through the linear interpolation method to evaluate the accuracy with which linear interpolation approximates the fully ray-traced spectra. The biggest error caused by the linear interpolation arises for spectra with a high spin BH but low accretion rate. Those spectra are the most sensitive to spin and accretion rate, implying large changes in X-ray flux in response to small parameter changes. As we can see from the lower panel of the left plot, the error in the interpolated luminosity density is $<10\%$ for all the cases considered over the $0.2-3$ keV energy band. If we only consider the energy range over which the luminosity density decreases by a factor of $<10^4$, the error is $<5\%$.    

The right hand side of Figure \ref{fig:er1} shows the spectra and the fractional error caused by interpolation for disks with high accretion rates. As we can see from the spectra, the luminosity density decreases as the disk inclination becomes more edge-on. This is because the edge of the disk blocks X-rays emitted by the inner region of the disk, preventing them from reaching the observer (W20). The error can be up to $40\%$ for an edge-on inclination. This effect was not seen in the left hand side panel of Figure \ref{fig:er1}, because of the smaller disk scale height at low Eddington ratio.  
Making the distance between grid points for the inclination two times smaller
only reduces the error on the luminosity determination marginally (it is still around $40\%$). In the next section, we test how well high-inclination sources can be fit using the library of model spectra, taking ASASSN-15oi as a test case. We note that the discrepancy between the detailed model calculations and the linear interpolation calculations at high inclinations becomes smaller for larger $M_\bullet$ 
($> 5.5\times 10^6 M_\odot$).

\section{Fitting Examples: ASASSN-14\lowercase{li} and -15\lowercase{oi}}
\label{asassns}

We have examined the flux errors caused by interpolating between library spectra and demonstrated
that the procedure is generally accurate relative to fitting on the fly with our W21 ray-tracing model.
In this section, we explore 
the effect on the $M_\bullet$ and $a_\bullet$ constraints from interpolating the library.
We consider two cases: the TDE ASASSN-14li and the likely edge-on TDE ASASSN-15oi. We fit their X-ray spectra using both the library and the full W21 ray-tracing approach and compare the results for $M_\bullet$ and $a_\bullet$.

In addition, we explore the constraints on parameters that were not investigated comprehensively in W20 ($\dot m_i$, $\theta$, and $N_{\rm H}$), because such a high-dimensional parameter sweep was computationally prohibitive using on-the-fly ray-tracing calculations. Now linear interpolation over the library of tabulated model spectra makes broader parameter surveys feasible. The fitting is done using XSPEC \citep{Arnaud1996}. 

For ASASSN-14li, we fit the 10 spectra with a slim disk plus power law model, where the power law component, {\sc po}, is only added to the fit function to describe the hard component detected during epoch 6.  For the slim disk component, we consider both the on-the-fly ray-tracing model of W21 and the library of pre-tabulated spectra from that model ({\sc slimd}). 
We fit the spectra of the 10 epochs simultaneously, requiring the inclination $\theta$, $M_\bullet$, and $a_\bullet$ to be the same for all epochs (the priors are the same as listed in Table \ref{tab:prior}), but allowing the absorption ({\sc phabs}) and accretion rate to vary between epochs.

\begin{figure}[ht!]
\plotone{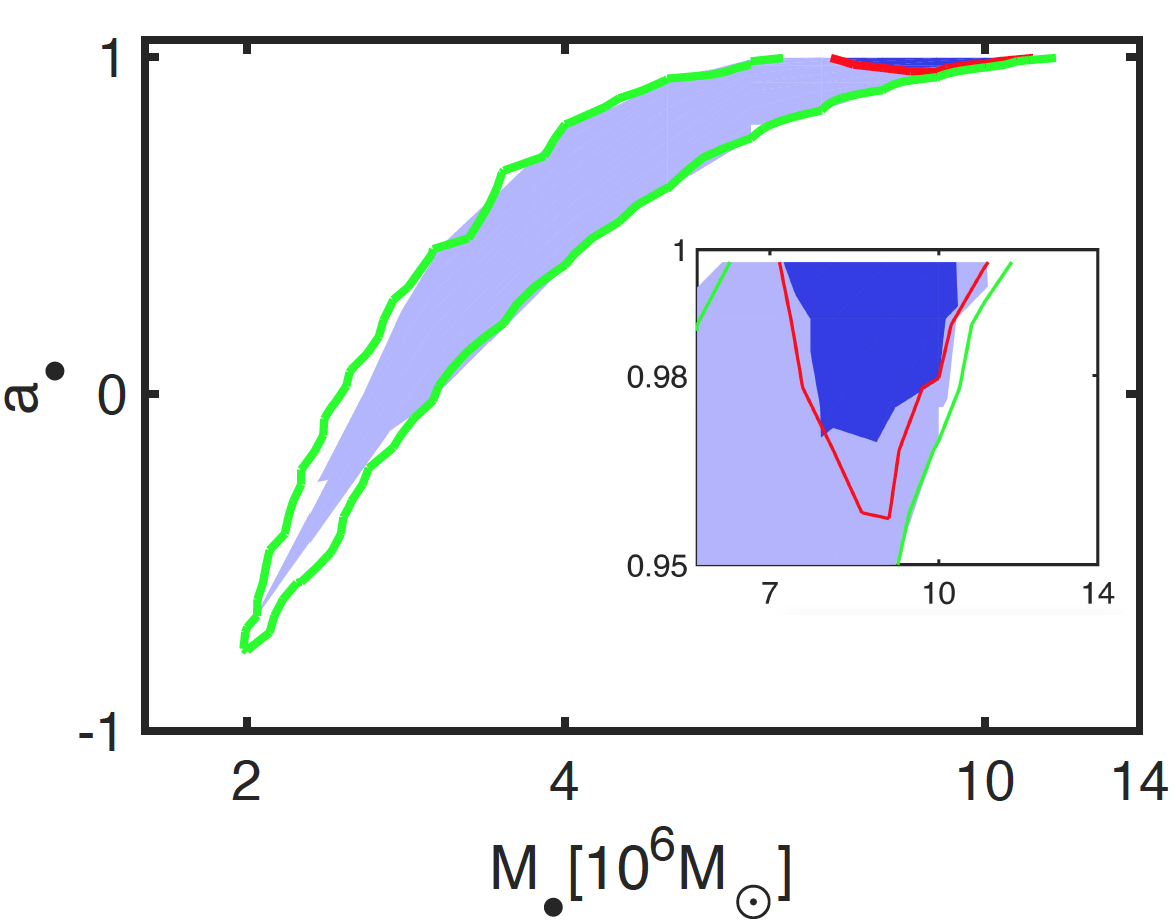}
\caption{Constraints on $M_\bullet$ and $a_\bullet$ for ASASSN-14li from both our original on-the-fly ray-tracing model (W21) and the library of model spectra. The blue and light blue regions represent the $1\sigma$ and $2\sigma$ contours for fits from the full W21 model, while the red and green lines are the $1\sigma$ and $2\sigma$ contours for fits from interpolating over the tabulated library. The close overlap between the contours indicates that using the library of model spectra approximates the results of W21 well for $M_\bullet$ and $a_\bullet$.} 
\label{fig:14lima}
\end{figure}

Figure \ref{fig:14lima} compares the constraints on $M_\bullet$ and $a_\bullet$
obtained through fitting the observed spectra with our original on-the-fly ray-tracing model (W21) and with the model spectral library. In both analyses, we fit the same observed spectra assuming the same parameter priors. Both the $1\sigma$ and $2\sigma$ contours overlap closely, indicating that the results
from the library and ray-tracing fits agree.
We have achieved this strong agreement despite a dramatic reduction in computational cost: generating the confidence contours in Figure \ref{fig:14lima} takes $\sim 10^4$ CPU hours for the full, on-the-fly ray-tracing calculation, but only $\sim 10$ CPU hours when interpolating over our pre-tabulated model library. 

For completeness, we also compare the $M_\bullet$ and $a_\bullet$ constraints from the ray-tracing models of W21 and W20 (which differ only by inclusion of angular momentum losses through radiation in W21). The two models give consistent results for the $2\sigma$ CLs, and W21 produces tighter constraints
for the $1\sigma$ CL (see Appendix \ref{app:14lima}).

\begin{figure*}[ht!]
\plotone{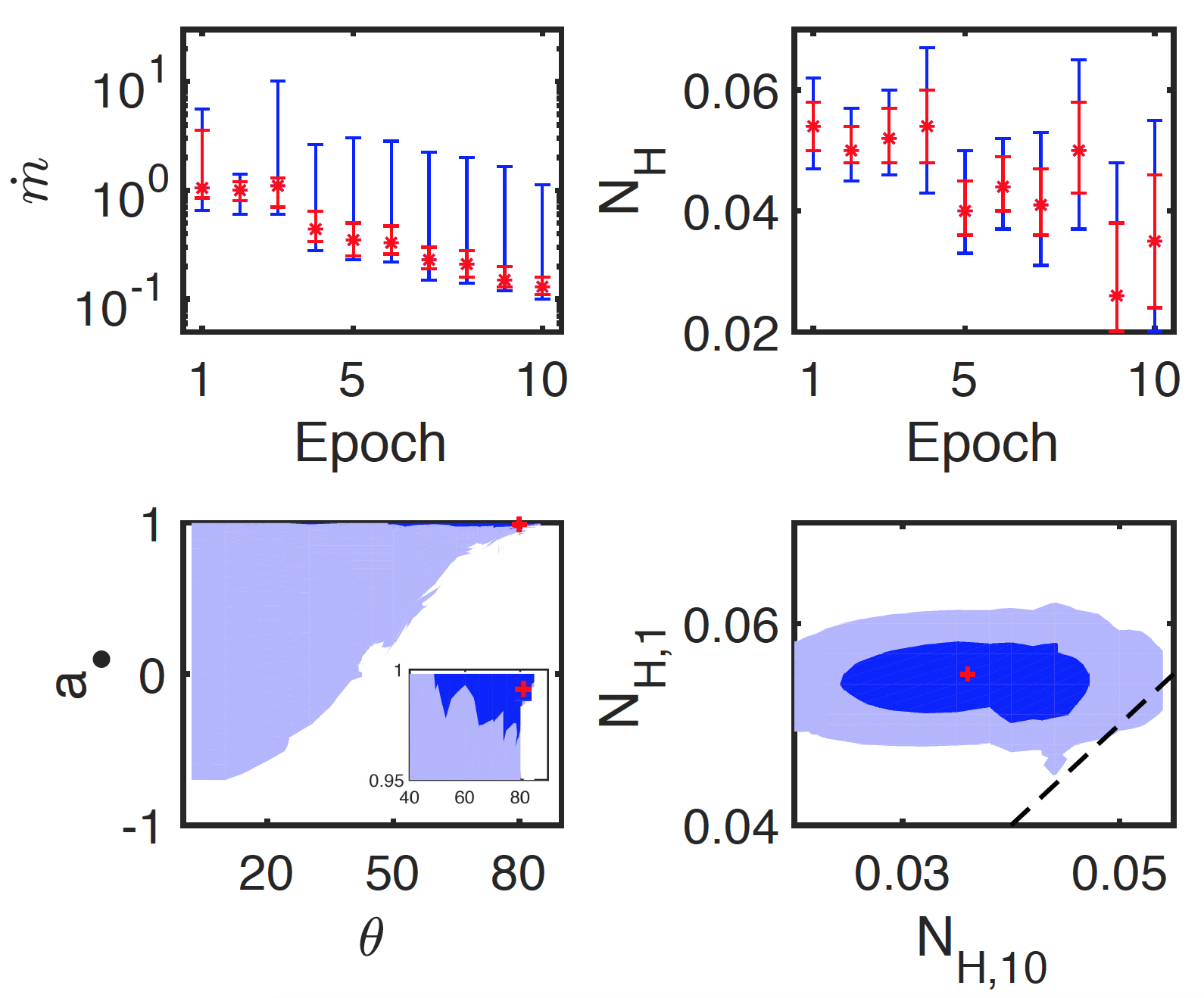}
\caption{Constraints on slim disk fit parameters for ASASSN-14li
from the model spectral library. The upper two panels show the constraints on the parameters $\dot m_i$ and $N_{\rm H, i}$ (${\rm 10^{22} cm^{-2}}$) for each epoch, derived, for the first time, allowing $M_\bullet$ and $a_\bullet$ to float. The $1\sigma$ and $2\sigma$ error bars are shown in red and blue, respectively. The lower left panel shows the constraints on $a_\bullet$ and inclination $\theta$ in degrees. The red cross shows the best fit. The blue and light blue regions represent the $1\sigma$ and $2\sigma$ contours, respectively. The inset zooms in on the best fit region. $\theta$ is poorly constrained at $2\sigma$ CL due to the large range in allowed \{$a_\bullet$,$M_\bullet$\} values.
The lower right panel shows the contours of $N_{\rm H,1}$ versus $N_{\rm H,10}$ compared to the $N_{\rm H,1} = N_{\rm H,10}$ line (dashed), demonstrating that $N_{\rm H}$ decreases in time.}
\label{fig:14li1}
\end{figure*}

Figure \ref{fig:14li1} shows the constraints on additional slim disk fit parameters
for ASASSN-14li
obtained from using the library of model spectra. Due to the resulting computational efficiency,
we can derive, for the first time,
constraints on $\dot m_i$, $N_{\rm H, i}$ (in units of $10^{22}$  $\rm cm^{-2}$), and $\theta$ for each epoch while allowing $M_\bullet$ and $a_\bullet$ to float.
The upper left panel shows the best-fit values and error bars for $\dot m_i$. We obtain a poor constraint (at $2\sigma$ CL) on the dimensionless mass accretion rate for most epochs, because the individual constraints on $M_\bullet$ and $a_\bullet$ are weak.
As a result, spectra with different mass accretion rates and 
\{$a_\bullet$,$M_\bullet$\} combinations can fit the data equally well.

The upper right panel shows the constraints on $N_{\rm H}$ for all epochs. Here, we constrain $N_{\rm H}$ well at most epochs, and we find a decline in $N_{\rm H}$ from epoch 1 to epoch 10. The fitted values of $N_{\rm H}$ in epochs 1 through 4 are larger than the value for the host galaxy plus the Galactic absorption combined ($4\times 10^{20} \rm{cm}^{-2}$; \citealt{Miller+15}) at a $2\sigma$ CL, while the fitted $N_{\rm H}$ in epochs 5 through 10 are consistent with it at $1\sigma$ CL. We also plot the contours of $N_{\rm H,1}$ versus $N_{\rm H,10}$ in the lower right panel. From these contours, we find a decrease in $N_{\rm H}$ from epoch 1 to epoch 10 with a $2\sigma$ CL. This indicates that the additional gas liberated in the TDE might be responsible for the enhanced absorption at early times, and that it disappears at late times. 

The lower left panel shows the joint constraint on $a_\bullet$ and $\theta$. The $1\sigma$ contour is small, because we constrain the spin to be $a_\bullet > 0.95$. $\theta$ is poorly constrained at $2\sigma$ CL due to the large range in allowed \{$a_\bullet$,$M_\bullet$\} values.

For the case of ASSASN-15oi, there are only two epochs of observations. We fit those two  spectra with the slim disk plus a  broken power law function ({\sc bknpow} in XSPEC). For the slim disk component, as for ASSASN-14li,
we use both the full on-the-fly ray-tracing 
model (W21) and the library of model spectra ({\sc slimd}). The {\sc bknpow} fit is used to describe the spectrum of the background photons; the {\sc bknpow} parameter values are determined through separate fitting of the background spectra. During the fitting of the two source spectra, the parameters of the broken power-law are kept fixed at their best-fit values. We note that the normalization of the broken power-law is allowed to float (this is consistent with the procedure employed in W20).  We use the floating normalization of the broken power-law to account for the difference in area over which the background and source photons are extracted.

In W20, we considered a variety of scenarios to explain the unusual time evolution \citep{Gezari+17} of ASASSN-15oi. Here we focus only on the ``slimming disk'' scenario of W20, so as to investigate the accuracy of parameter estimation with the library in the special case of  nearly edge-on TDEs. In W20, 
we used the mass fallback law in \citet{GuillochonRamirezRuiz13} as a prior for this scenario.
However, for our re-analysis using the library of model spectra, it is computationally feasible to keep $\dot m$ as a free parameter.
To remain close to the other priors used in W20, we allow $\dot{m}$ to float, but set 
$20 < \dot m_1 < 600$ Edd and 
$70^\circ < \theta <90^\circ$. With these assumptions, we fit the spectra with the ray-tracing model of W21 and the library of model spectra. 

\begin{figure}[ht!]
\plotone{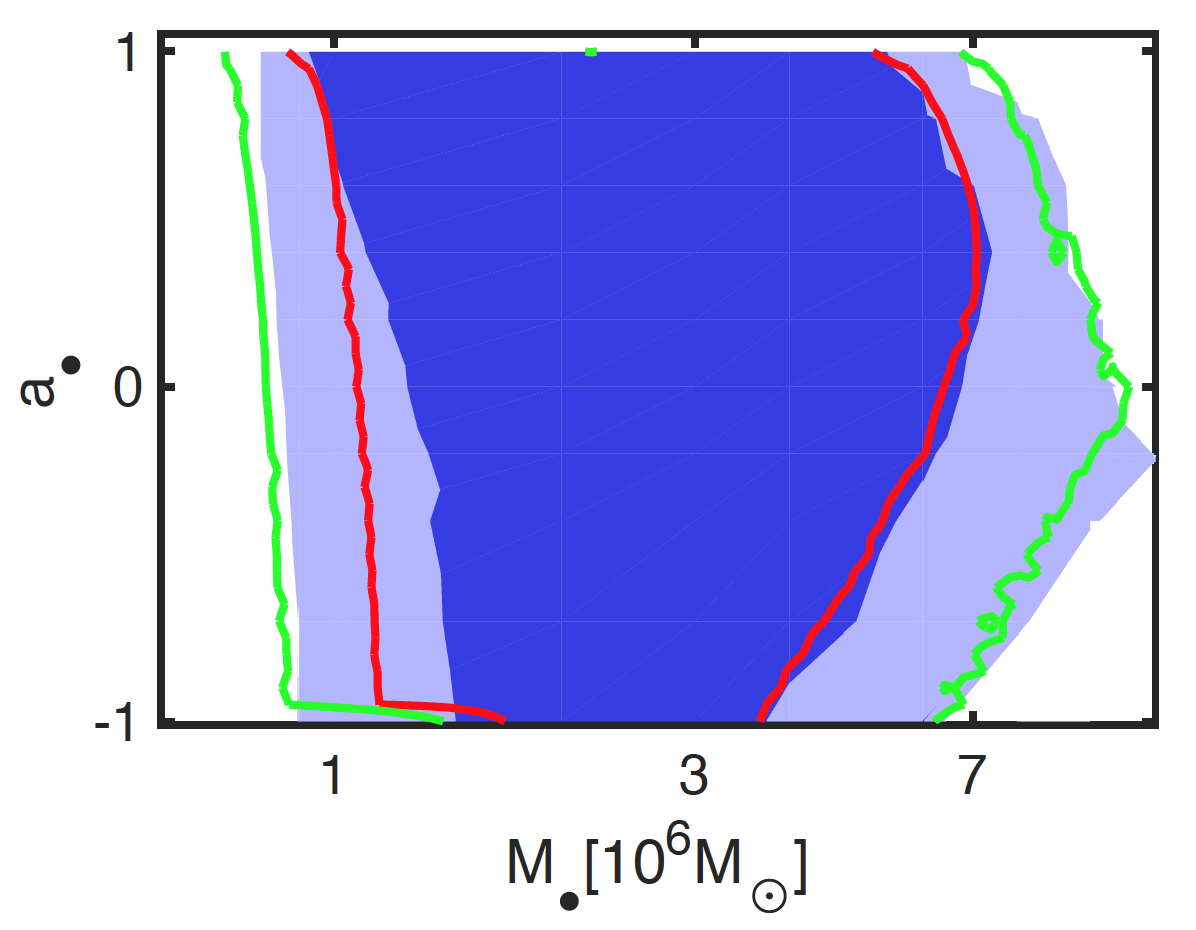}
\caption{Constraints on $M_\bullet$ and $a_\bullet$ for ASASSN-15oi from fitting the two epochs of observed X-ray spectra with the original full ray-tracing procedure of W21 and with the library of model spectra. The blue and light blue regions denote the $1\sigma$ and $2\sigma$ contours resulting from the fits using the on-the-fly ray-tracing model, while the red solid and green solid lines are the $1\sigma$ and $2\sigma$ contours using the model spectral library. The contours overlap well, indicating that, for this edge-on disk case, the library of model spectra also approximates the results of the original full ray-tracing model of W21 well on $M_\bullet$ and $a_\bullet$ estimation. 
}
\label{fig:15oima}
\end{figure}

Figure \ref{fig:15oima} compares the constraints on $M_\bullet$ and $a_\bullet$ obtained through fits using the library of model spectra versus on-the-fly ray-tracing (W21). The $1\sigma$ and $2\sigma$ constraints from the two approaches overlap, demonstrating the benefit of fitting faster with the pre-tabulated library.

\begin{figure}[ht!]
\gridline{\fig{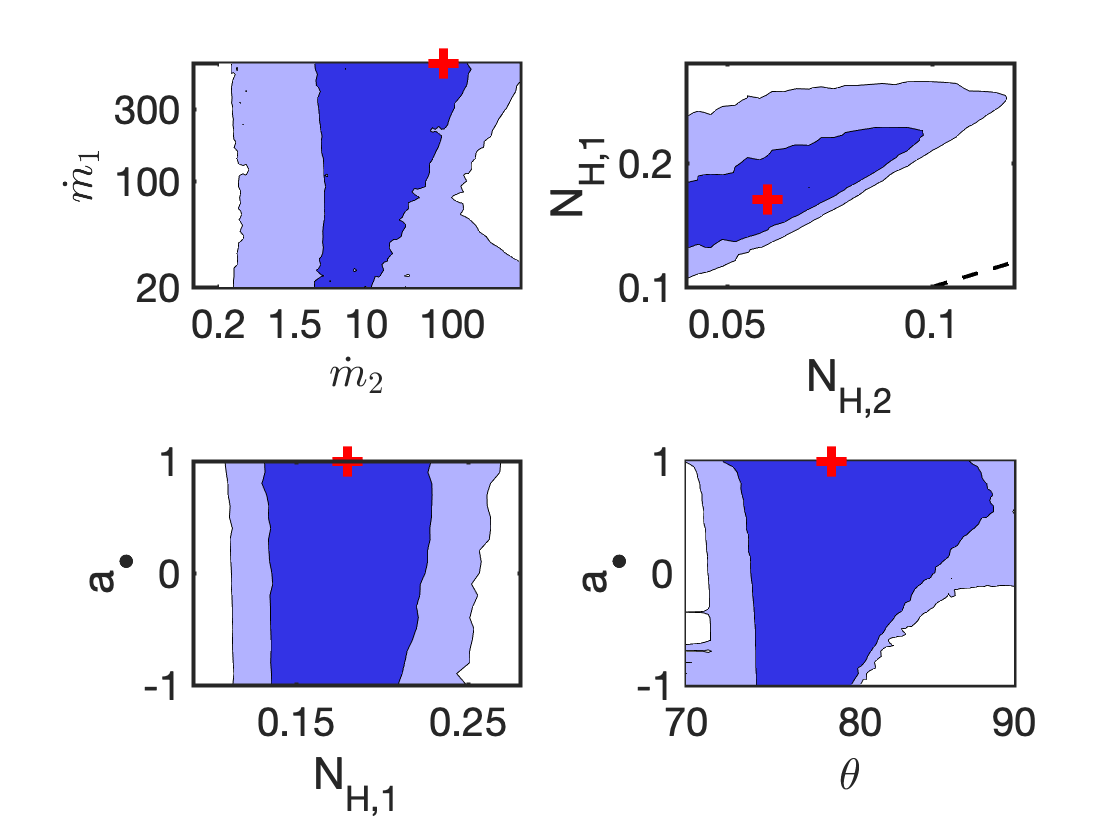}{0.5\textwidth}{}}
\caption{Constraints on slim disk fit parameters for ASASSN-15oi from the library of model spectra, where
$M_\bullet$ and $a_\bullet$ are allowed to float for the first time.
The blue and light blue regions denote the $1\sigma$ and $2\sigma$ contours, respectively, that we find by linearly interpolating over the library.
The red cross shows the best fit. N$_{H,i}$ is given in units of $10^{22}$ $\rm{cm^{-2}}$, and $\theta$ is in degrees. The accretion rate is only loosely constrained, and the spin is unconstrained. The best fit $\theta$ and its 1$\sigma$ contours lie well within the assumed prior range for $\theta$; this strong constraint indicates that the disk is nearly edge-on. There are tight constraints on $N_{\rm H}$ for both epochs. We confirm (at $>3\sigma$ confidence) the decrease in $N_{\rm H}$ in time seen by W20 (dashed line denotes $N_{\rm H,1}=N_{\rm H,2}$). This behavior may arise from the source being observed at high inclination \citep{LuBonnerot19}.}
\label{fig:15oip}
\end{figure}

Figure \ref{fig:15oip} shows the constraints on additional slim disk model parameters
for ASASSN-15oi that we obtain from fitting the library model spectra to the two epochs of XMM-{\it Newton} data. As we did for ASASSN-14li, we constrain these parameters for the first time while allowing 
$M_\bullet$ and $a_\bullet$ to float. The best-fit values of the accretion rate are only loosely constrained, and $a_\bullet$ is unconstrained. The constraint on $\theta$, which is well within the assumed prior range for $\theta$, indicates that the disk is in fact nearly edge-on. The poor constraint on spin may due to the X-ray photons from the inner part of the edge-on disk being blocked during the early epoch. For both epochs, there are strong constraints on $N_{\rm H}$. We find, at $>3\sigma$ confidence, the decrease in $N_{\rm H}$ over time also seen by W20; this behavior may be related to the source being viewed close to edge-on \citep{LuBonnerot19}.

Here we have refit the events ASASSN-14li and ASASSN-15oi with the library of model spectra. We successfully reproduce the constraints on $M_\bullet$ and $a_\bullet$ obtained from the more computationally costly method of on-the-fly ray-tracing. In addition, we obtain new constraints on the $\dot m$, $\theta$ and $N_{\rm H}$ parameters. This indicates that fits obtained by linearly interpolating over the library of model spectra are reliable representations of the results that would have been obtained had the models been calculated using the full ray-tracing code at all parameter values.

\section{Dependence of parameter constraints on observational data and disk properties}
\label{mock_fitting}

In this section, we investigate how the accuracy of estimated parameters---$M_\bullet$, $a_\bullet$, mass accretion rate, and disk inclination---is affected by the number of observing epochs and the number of X-ray photons detected. We also explore the impact of photoelectric absorption by including model {\sc phabs}.  In addition, we survey the parameter space to determine which types of disks yield stronger or weaker constraints on $M_\bullet$ and $a_\bullet$. For instance, holding all else equal, does a highly super-Eddington mass accretion rate allow for a stronger or weaker constraint on $M_\bullet$ and $a_\bullet$?

We use the model {\sc slimd}
and the 
XMM-{\it Newton} pn response file
to generate mock observational spectra.
These mock spectra have no background component, but they include statistical noise to better simulate the observations.
The parameters used to generate the mock spectra are given in Table \ref{tab:mock}. We rebin the data such that a bin contains at least one X-ray photon (i.e., count). We use the library of model spectra to fit these mock spectra, employing Poisson statistics \citep{Cash1979}.

We first investigate if there are parameter degeneracies using a single-epoch mock spectrum. Then, we explore the effect of the total number of counts. Moreover, we show that multi-epoch spectra help to break degeneracies between parameters. We also demonstrate that the early-time epochs of edge-on TDEs have important constraining power. 

\begin{deluxetable*}{ccccccccccccccc}
\tablecaption{Input Parameters and Photon Number Counts for Mock Observational Spectra}
\tablewidth{0pt}
\tablehead{
\colhead{Spectrum }&&\colhead{Counts}&&\colhead{$N_{\rm H} [10^{22} \rm{cm}^{-2}]$ }&&\colhead{$\dot m$ [Edd]}&&\colhead{ $\theta$ [$^\circ$]}&&\colhead{$M_\bullet$ [$M_\odot$]}&&\colhead{$a_\bullet$} &&\colhead{Exp time [ks]} }
\startdata
1 && 46393 && 0.03 && 3 && 45 && $10^6$ && 0.9&& 3\\
2 && 46676 && 0.03 && 3 && ... && $10^6$ && 0&& 16\\
3 && 46022 && 0.03 && 3 && ... && $10^4$ && 0.9&& 55\\
4 && 47697 && 0.03 && 0.2 && ... && ... && ...&& 200\\
\hline
5 && 49621 && 0.03 && 1 && 45 && $10^6$ && 0.9&& 9\\
6 && 47981 && 0.03 && 0.9 && ... && ... && ...&& 10\\
7 && 48328 && 0.03 && 0.8 && ... && ... && ...&& 12\\
8 && 45978 && 0.03 && 0.7 && ... && ... && ...&& 14\\
9 && 48466 && 0.03 && 0.6 && ... && ... && ...&& 19\\
10 && 46089 && 0.03 && 0.5 && ... && ... && ...&& 25\\
11 && 47314 && 0.03 && 0.4 && ... && ... && ...&& 40\\
12 && 47827 && 0.03 && 0.3 && ... && ... && ...&& 75\\
13 && 192620 && 0.03 && 1 && ... && ... && ...&& 35\\
\hline
14 && 18987 && 0.03 && 1 && 45 && $10^4$ && 0.9&& 50\\
15 && 17242 && 0.03 && 0.9 && ... && ... && ...&& 50\\
16 && 15500 && 0.03 && 0.8 && ... && ... && ...&& 50\\
17 && 13533 && 0.03 && 0.7 && ... && ... && ...&& 50\\
\hline
18 && 3032 && 0.03 && 10 && 80 && $10^6$ && 0.9&& 2\\
19 && 20769 && 0.03 && 2 && ... && ... && ...&& 4.2\\
20 && 21342 && 0.03 && 1 && ... && ... && ...&& 9\\
21 && 20684 && 0.03 && 0.9 && ... && ... && ...&& 10\\
22 && 21085 && 0.03 && 0.8 && ... && ... && ...&& 12\\
23 && 20134 && 0.03 && 0.7 && ... && ... && ...&& 14\\
\enddata
\label{tab:mock}
\tablecomments{We assume an absorbed slim disk model, {\sc phabs*slimd}, to generate the mock spectra using the XMM-{\it Newton} pn response file. We adopt $D_L=100~\rm{Mpc}$ (corresponding to $z=0.023$) for all spectra, and we use {\sc grppha} to rebin the data so that there is at least one count per spectral energy bin.}
\end{deluxetable*}

\subsection{Parameter degeneracies}
\label{sec:pd}

\begin{figure*}[ht!]
\gridline{\fig{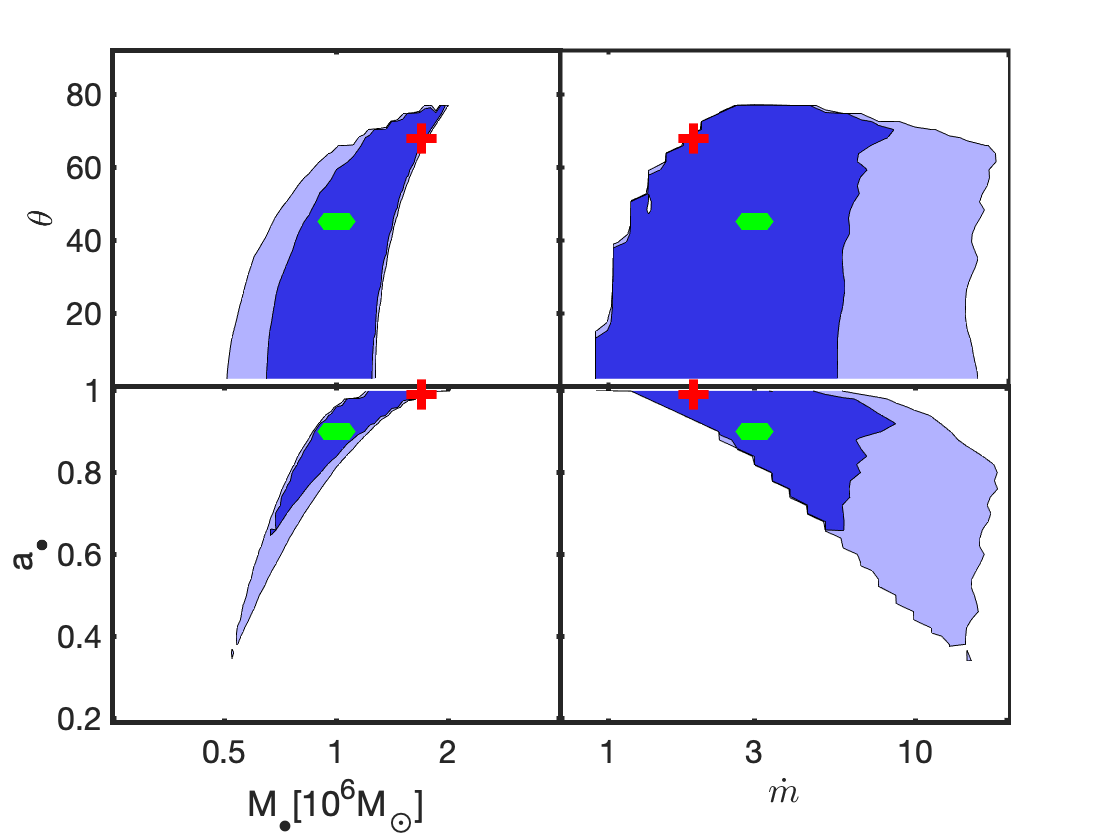}{0.45\textwidth}{$\dot m=3$, $a_\bullet=0.9$, and $M_\bullet=10^6 M_\odot$}
          \fig{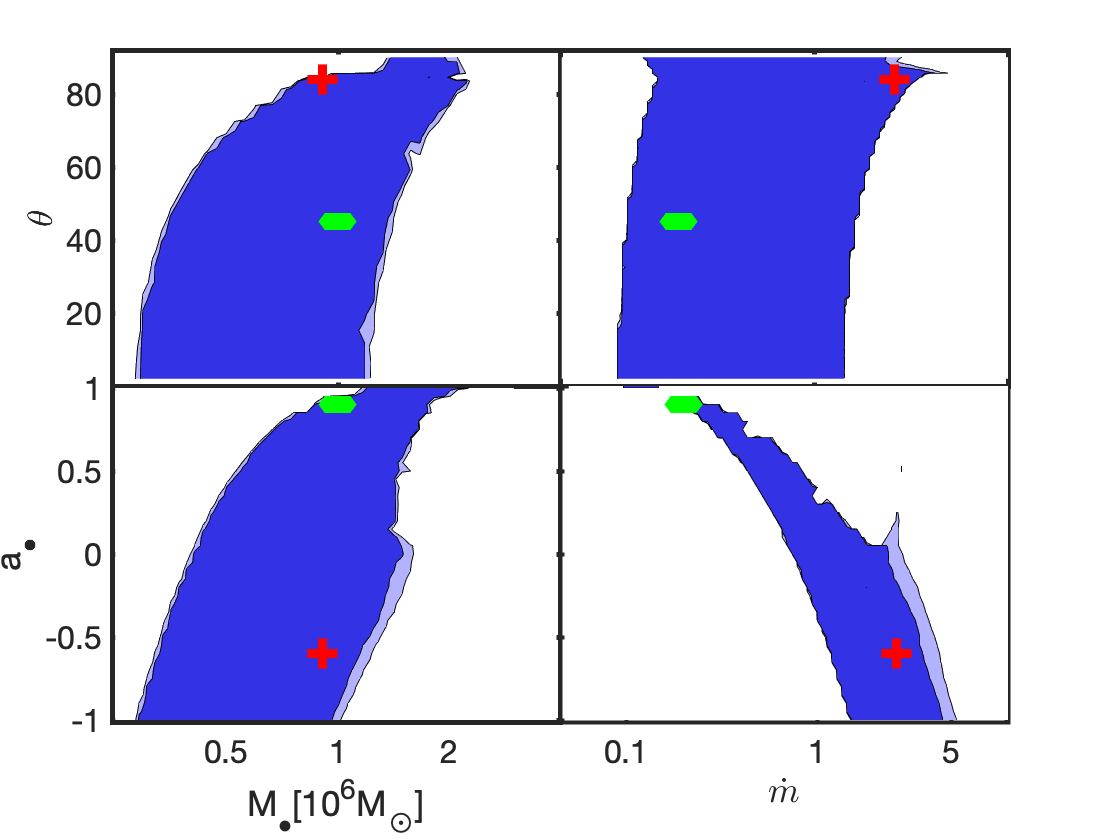}{0.45\textwidth}{$\dot m=0.2$, $a_\bullet=0.9$, and $M_\bullet=10^6 M_\odot$}
}
\gridline{\fig{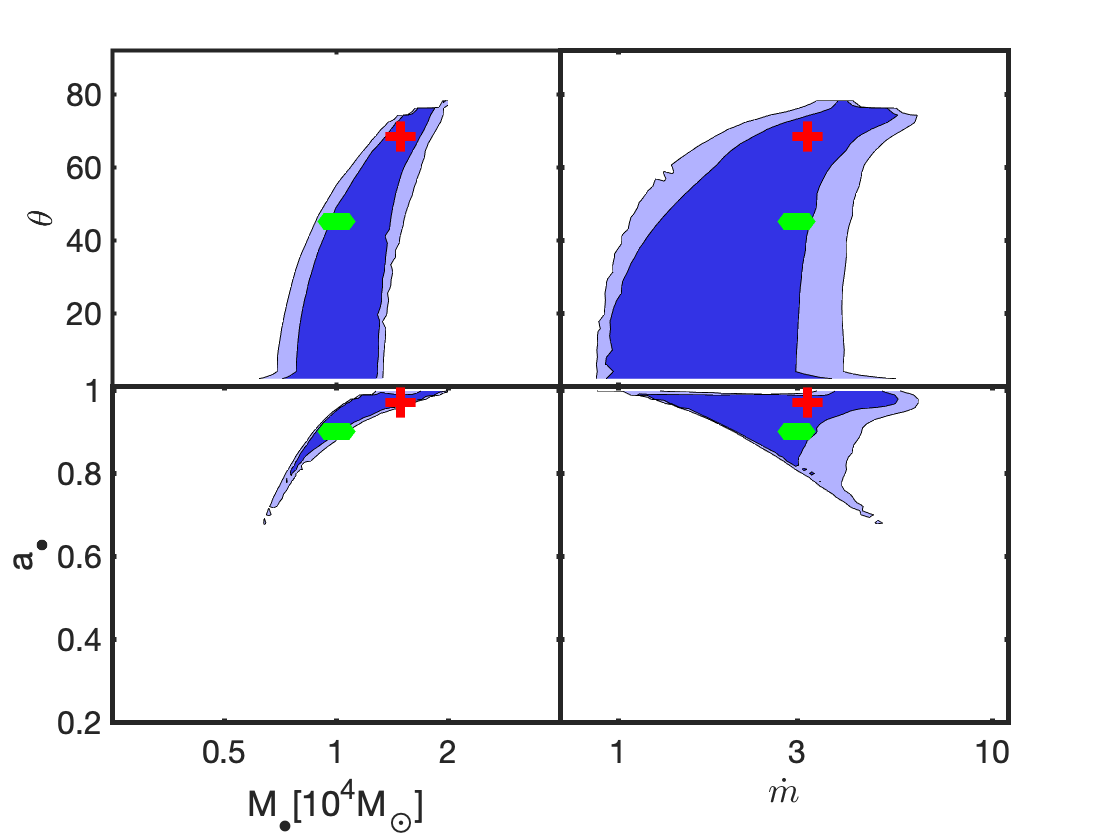}{0.45\textwidth}{$\dot m=3$, $a_\bullet=0.9$, and $M_\bullet=10^4 M_\odot$}
\fig{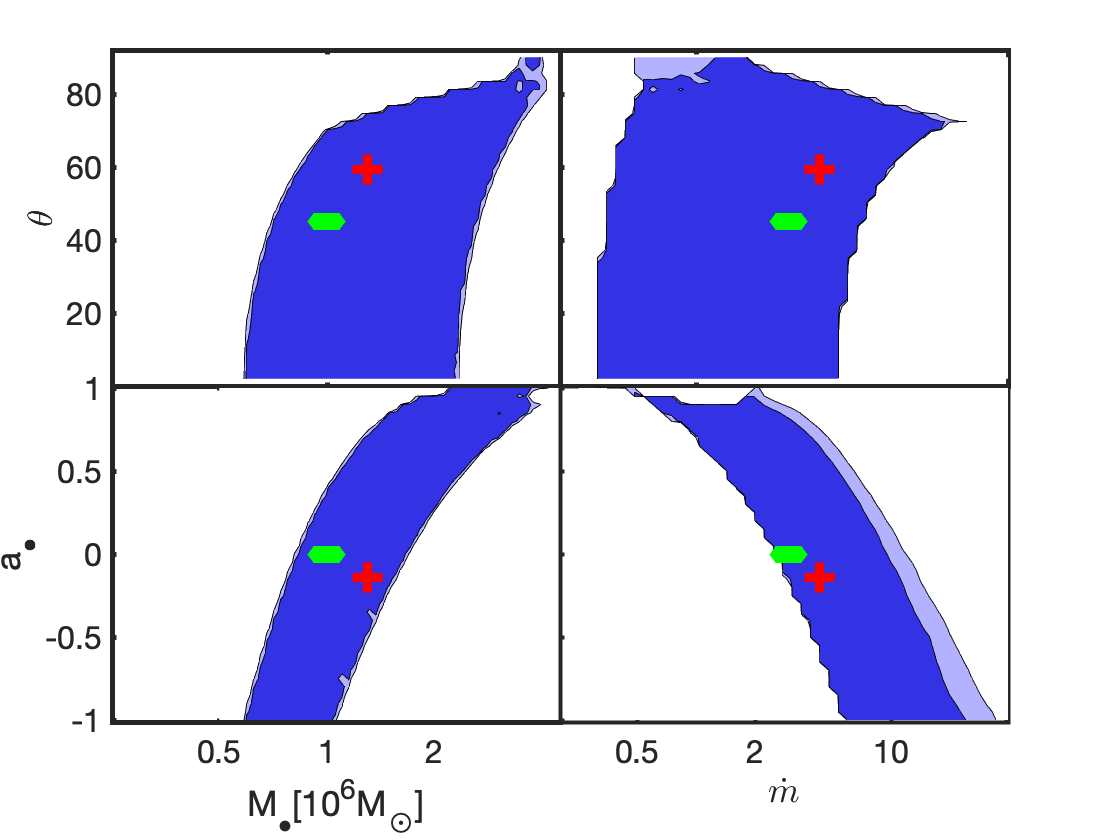}{0.45\textwidth}{$\dot m=3$, $a_\bullet=0.0$, and $M_\bullet=10^6 M_\odot$}
}
\caption{Contours of different parameters from fitting the mock observed spectra with the library of model spectra. The blue and light blue regions denote the $1\sigma$ and $2\sigma$ contours, respectively.
The green diamond is positioned at the ``true'' parameter values (i.e., the values used to generate the mock spectra). The red cross denotes the best-fit value of each parameter pair. These plots show that the best-fit values of accretion disk parameters are often poorly constrained (degenerate) for these single-epoch spectral fits. Single-epoch spectral fits can provide constraints on $M_\bullet$. For high BH spin and super-Eddington spectra, joint constraints on $a_\bullet$ and $M_\bullet$ are narrow, and the situation improves for lower mass BHs, such as an IMBH. While the $M_\bullet$ and $a_\bullet$ contours become significantly smaller, the $1\sigma$ error for $M_\bullet$ only slightly shrinks due to the
$\{M_\bullet, a_\bullet\}$
degeneracy.
}
\label{fig:ma}
\end{figure*}

Figure \ref{fig:ma} shows the best-fit confidence contours for different parameters derived from library fits to different mock observed spectra. Here, we consider four mock spectra with different combinations of $\dot m$, $a_\bullet$ and $M_\bullet$, sharing a common source inclination of $\theta=45^\circ$\footnote{ Because inclination is poorly constrained and strongly degenerate with other parameters when the disk is not edge-on, we choose $45^\circ$ to test this case here; we explore an edge-on case ($\theta = 80^\circ$) more like ASASSN-15oi in \S{\ref{edg-on}}.}.
The input parameters for these mock spectra can be found in the first four rows of Table \ref{tab:mock} and are shown under each of the figure panels. For each mock spectrum, we adjust the exposure time so that all have a similar number of counts in the $0.3-10$~keV band. 

The four top left panels show the fits to a super-Eddington, high-$a_\bullet$ spin, and relatively high-$M_\bullet$ source spectrum. The $M_\bullet$ and $a_\bullet$ contours show that
$M_\bullet$ and $a_\bullet$ are well constrained, indicating that the spectral shape is sensitive to these two parameters. From the other contours, we find that $\dot m$ and $\theta$ are constrained poorly, and that both are highly degenerate with other disk parameters. The $2\sigma$ range of $\dot m$ extends to 
$\dot m$ 
values larger than 10 (in Eddington units), due to the fact that the X-ray luminosity is weakly sensitive to accretion rate for highly super-Eddington slim disks (\citealt{Abramowicz+88}; W20). The confidence contours of $\theta$ extend up to $\sim 80^\circ$, but not higher, because an even higher inclination would cause the accretion disk to block most X-rays coming from the inner disk, reducing the observable flux. 

The four top right panels show results for the same input parameters 
as the four top left panels
except for accretion rate, which has now been set to the sub-Eddington value of $\dot m=0.2$. From the $M_\bullet$ and $a_\bullet$ contours, we see that the model fits still constrain $M_\bullet$, but no longer 
$a_\bullet$. The low accretion rate makes the disk cooler, moving the spectrum to lower X-ray energies, which weakens the constraint on $a_\bullet$. The constraints on $\dot m$ and $\theta$ become even weaker when compared with the case of the super-Eddington spectrum. In order to constrain $a_\bullet$ using a single epoch of data when the mass accretion rate is sub-Eddington, one needs to know the value of $M_\bullet$ so that it can be held fixed in the fit. 

The four bottom right panels show the results for a mildly super-Eddington 
disk with $a_\bullet=0$. The constraints on all parameters are weaker than in the high-spin case shown in the upper left panel. Similar to the sub-Eddington mass accretion case, a lower spin makes the disk cooler, resulting in a softer spectrum and poorer parameter estimates. 

The four bottom left panels show the fit results for an intermediate mass black hole (IMBH): $M_\bullet$ is reduced to $10^4$~M$_\odot$, with $\dot{m}=3$ and $a_\bullet=0.9$ as in the top left panel. The confidence contours on each parameter are the tightest for all of the four cases shown in this figure. 
This is because a smaller $M_\bullet$ results in a hotter disk (roughly, $T_e \propto M_\bullet^{-0.25}$), meaning that a larger portion of the source spectrum falls in the X-ray band. We note that for such a low $M_\bullet$, the Eddington luminosity is 100 times lower than
for the other three panels (where $M_\bullet=10^6$~M$_\odot$), implying that the system has to be $\approx$10 times\footnote{The factor will actually be less than 10, because an IMBH has a harder spectrum than a SMBH and thus
a larger fraction of the spectrum will fall in the X-ray band.} closer to yield the same number of X-ray counts for fixed observing time, as we have assumed in making these comparisons.

In all of these efforts to fit mock spectra with the library of model spectra, the input values of all parameters ($M_\bullet$, $a_\bullet$, $\dot{m}$, $\theta$) are recovered within the $1\sigma$ confidence contours. In all cases, we find that: 1) $M_\bullet$ can be constrained relatively well; 2) both $\dot m$ and $\theta$ are poorly constrained, as they are highly degenerate with other parameters. The spin $a_\bullet$ can be well-constrained for cases with a high BH spin, super-Eddington accretion rates, or a BH in the intermediate-mass regime (although, in the IMBH case, it is more observationally challenging to obtain a large number of counts).      
As a rough heuristic summarizing these results, we find stronger constraints on all parameters when the information-rich peak of the quasi-thermal spectrum lies in (or at least closer to) the observing band, and weaker constraints when the observing band only contains the information-poor Wien tail.

\subsection{Effect of X-ray photon counts}
\label{sec:counts}

In this section, we explore the dependence of our joint $\{M_\bullet, a_\bullet \}$ constraints on the total number of X-ray counts. In order to quantify this dependence, we use a figure of merit (FoM). 
Following the Dark Energy Task Force team \citep{Andreas+2006}, we define the FoM as the reciprocal of the area of the error ellipse in the $\{\log_{10}(M_\bullet), a_\bullet\}$ plane. The FoM can be calculated as
\begin{equation}
{\rm FoM}=\frac{1}{\sqrt{|{\rm Cov}(\log_{10}(M_\bullet), a_\bullet)|}},
\end{equation}
where ${\rm Cov}(\log_{10}(M_\bullet), a_\bullet)$ is the covariance matrix of $\log_{10}(M_\bullet)$ and $a_\bullet$. A larger (smaller) FoM indicates a better (weaker) constraint on $\log_{10}(M_\bullet)$ and $a_\bullet$.   

\begin{figure}[ht!]
\plotone{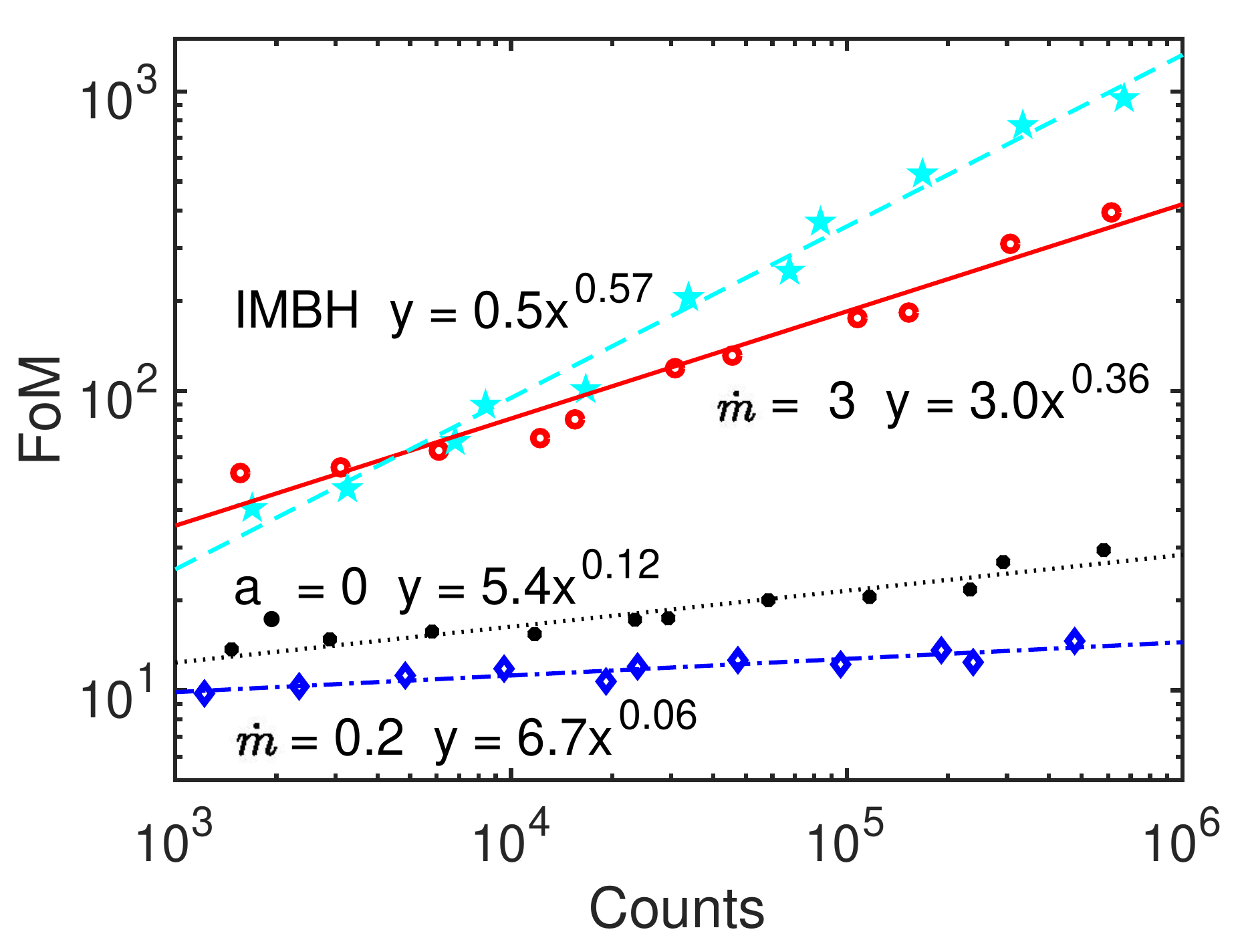}
\caption{Increase in the figure of merit (FoM) of a fit as a function of the 
number of counts in the spectrum. The data points denote the pairs (FoM, counts) found in fitting slim disk models to mock spectra from four different parameter sets: a super-Eddington IMBH ($M_\bullet = 10^4 M_\odot$, $\dot m=3$, cyan stars), a super-Eddington SMBH ($M_\bullet = 10^6 M_\odot$, $\dot m=3$, red circles), a sub-Eddington SMBH ($M_\bullet = 10^6 M_\odot$, $\dot m=0.2$, blue diamonds), and a low spin SMBH ($M_\bullet = 10^6 M_\odot$, $\dot m=3$, black circles). 
The cyan dashed, red solid, and blue dash-dotted lines show the best-fit power laws to the (FoM, Counts) pairs. This figure shows that, for all cases, the ability to constrain $\{M_\bullet, a_\bullet\}$ increases as a power law with increasing numbers of counts in the X-ray spectra.  The power law is steeper for parameters that yield higher-temperature disks, i.e., the quasi-thermal peak is closer to the X-ray observing band.}
\label{fig:fom}
\end{figure}

In order to explore the relationship between the FoM and the number of counts in the spectra, we consider four different sets of spectra: super-Eddington IMBH ($M_\bullet = 10^4 M_\odot$, $\dot m=3$), super-Eddington SMBH ($M_\bullet = 10^6 M_\odot$, $\dot m=3$), sub-Eddington SMBH ($M_\bullet = 10^6 M_\odot$, $\dot m=0.2$), and low spin SMBH ($M_\bullet = 10^6 M_\odot$, $\dot m=3$). We use the {\sc slimd} model to generate 11 mock spectra for each set, each with the same $a_\bullet=0.9$ (except for the low spin case)  and $\theta=45^\circ$.
Each set undergoes mock observations with a range of different exposure times\footnote{For the IMBH case, the exposure times are: 2, 4, 8, 10, 20, 40, 80, 100, 200, 400 and 800 ks.  For the super-Eddington SMBH case, the exposure times are: 0.1, 0.2, 0.4, 0.8, 1.0, 2.0, 4.0, 8.0, 10, 20, and 40 ks. For the sub-Eddington SMBH case, the exposure times are: 5, 10, 20, 40, 80, 100, 200, 400, 800, 1000, and 2000 ks. For the low spin SMBH case, the exposure times are: 0.5, 1, 2, 4, 8, 10, 20, 40, 80, 100, and 200 ks.} that are chosen for equal numbers of counts among the four sets. We fit these spectra using the {\sc slimd} model, allowing all the parameters to vary freely. We do not get the covariance matrix directly from the fitting result, as it is not accurately determined in {\sc XSPEC}, which only estimates the Fisher Matrix \citep{Arnaud1996}. However, we run the {\sc steppar} command for the parameters $M_\bullet$ and $a_\bullet$, and then calculate the covariance matrix with the likelihood function (derived from $Cstat$) and the corresponding value of $M_\bullet$ and $a_\bullet$. 

Figure~\ref{fig:fom} shows the FoM versus the number of X-ray counts for the four sets of spectra that we simulated. In all cases, a power law function fits the data well. The power-law index varies depending on the underlying parameters: it is highest for the IMBH case and lowest for the sub-Eddington SMBH case.  
Interestingly, the normalization of the IMBH case's power-law is lower than that of the super-Eddington SMBH case, so that the FoM of the IMBH only 
bests 
the super-Eddington SMBH once the number of counts exceeds several thousand.  
In summary, the FoM power-law indices fitted here quantify the trend noted heuristically in \S \ref{sec:pd}: hotter disks generally provide more constraining power per X-ray photon.

\subsection{Constraints from fits to multi-epoch data}

Unlike most active galactic nuclei (AGN), the accretion rate varies strongly with time in TDE disks. Therefore, multi-epoch X-ray spectroscopy has the potential to obtain more stringent constraints on the best-fit parameters for IMBH/SMBH accretion than it does for typical AGN. In this section, we explore how multi-epoch spectroscopy influences parameter estimation. 
We first examine the constraints obtained for a low or high accretion rate source. In order to do so, we generate nine epochs of mock spectroscopic data for SMBH TDEs, using {\sc phabs*slimd} as an input model (see rows 5 to 13 in Table~\ref{tab:mock}). 
Next, we study the case where multi-epoch spectra are obtained for IMBH TDEs. We create four mock spectra with an exposure time of $50~\rm{ks}$ (rows 14 to 17 in Table~\ref{tab:mock}).
In this analysis, when fitting the simulated spectra, we fix the value of $N_{H}$ as a prior and do not allow it to float or vary in time.

\begin{figure*}[ht!]
\gridline{\fig{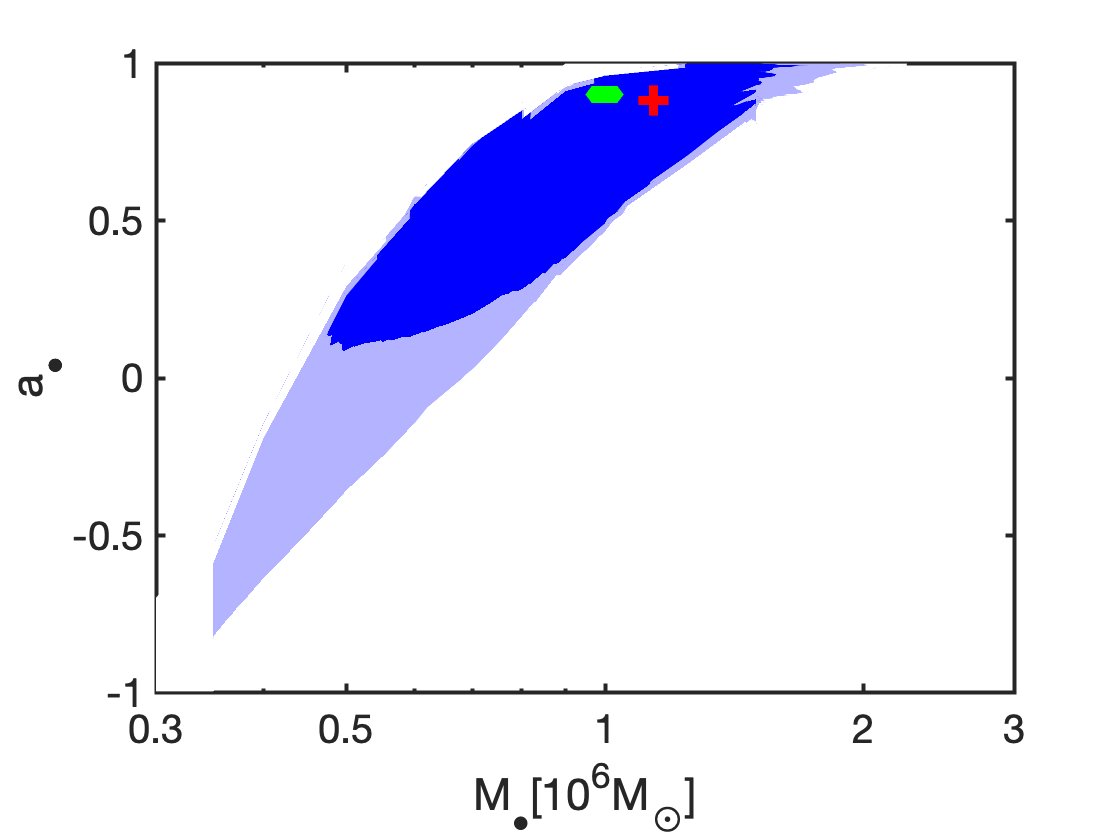}{0.45\textwidth}{$\dot m=$ 1, 0.9, 0.8, and 0.7.}
          \fig{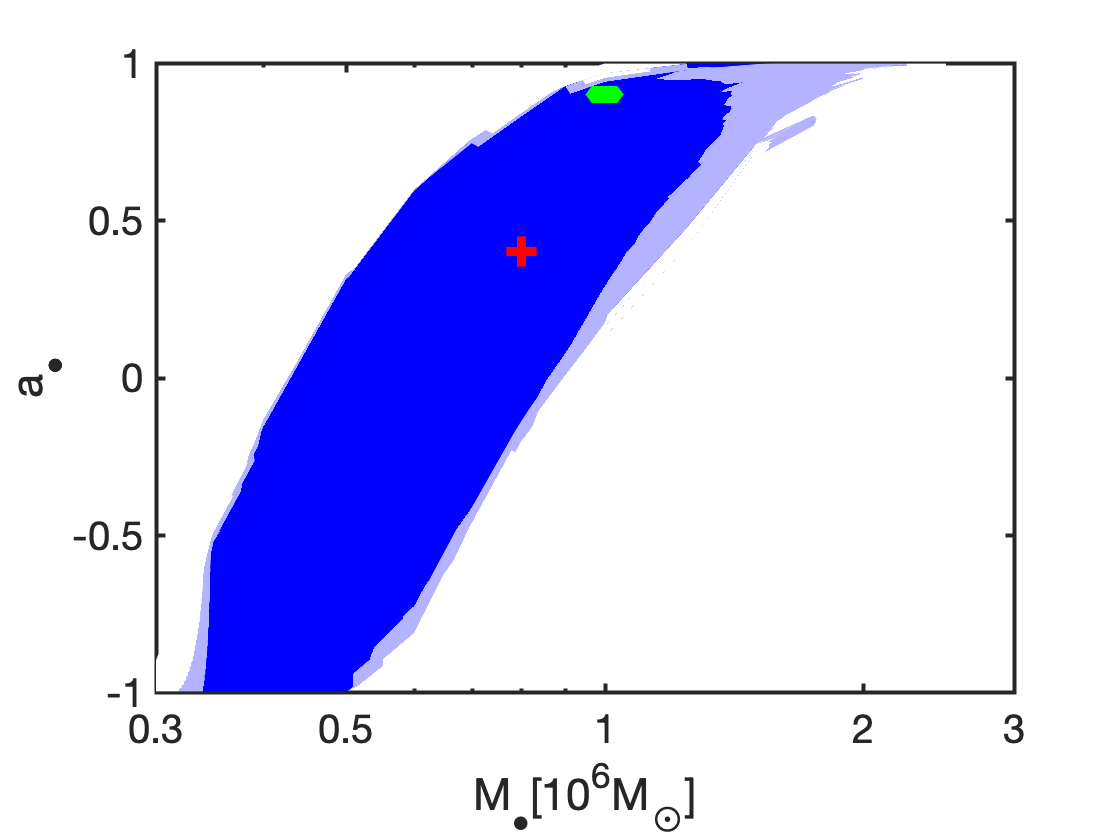}{0.45\textwidth}{$\dot m=$ 0.6, 0.5, 0.4, and 0.3.}
}
\gridline{ \fig{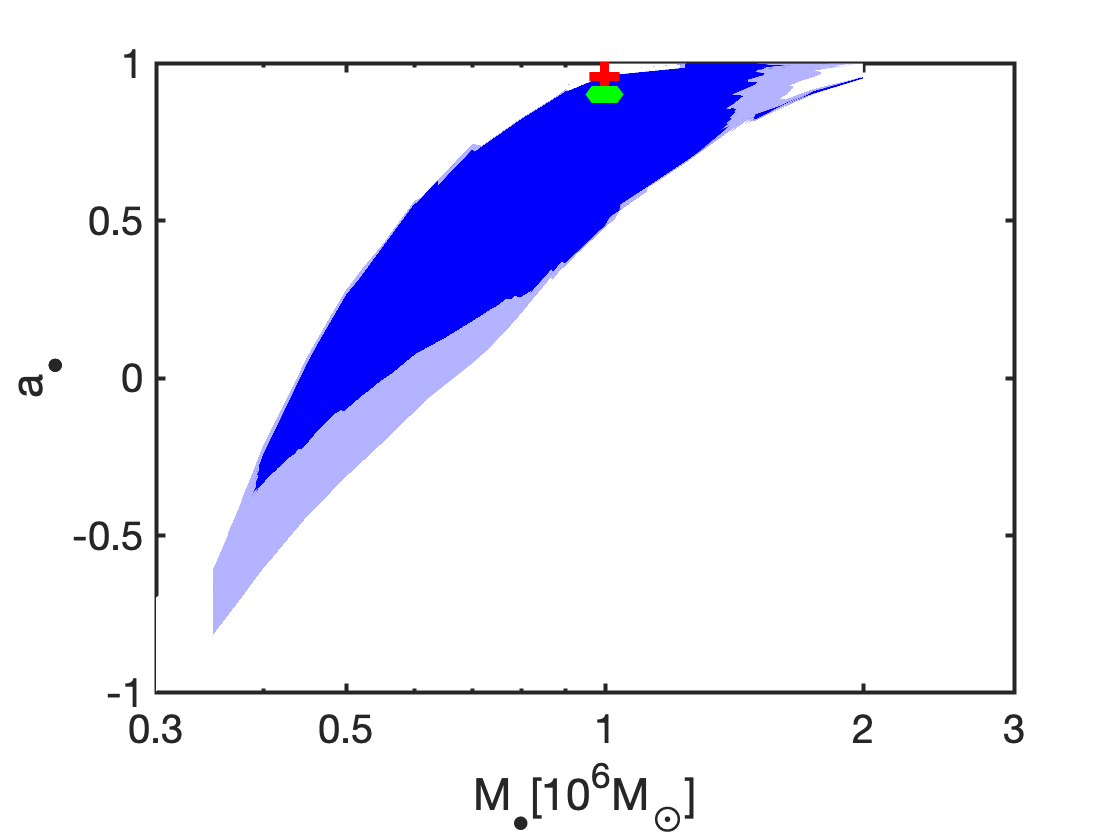}{0.45\textwidth}{$\dot m=1$ with counts = 192620.}
\fig{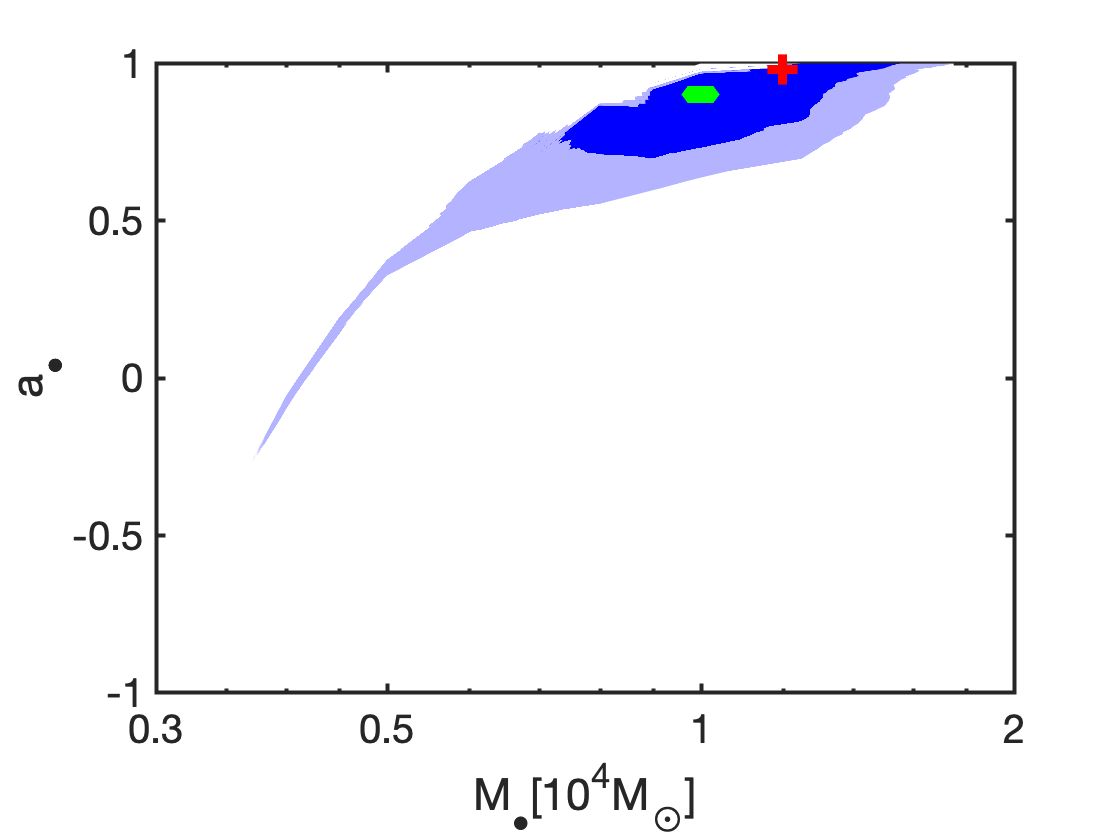}{0.45\textwidth}{$\dot m=$ 1, 0.9, 0.8, and 0.7.}
}
\caption{Constraints on $M_\bullet$ and $a_{\bullet}$ from simultaneous slim disk fits to multi-epoch mock spectra. The blue and light blue regions denote the 1$2\sigma$ and $2\sigma$ confidence contours, respectively. The green diamond is at the true value for the parameter pair, while the red cross is placed at the best-fit value.  The upper two panels show that multi-epoch spectra featuring higher Eddington ratios (top left panel; rows 5 to 8 in Table~\ref{tab:mock}) constrain both $M_\bullet$ and $a_\bullet$ better (FoM $= 39.1$) than 
when the mass accretion rate is sub-Eddington (top right panel; Table \ref{tab:mock} rows 9-12; FoM $= 19.0$). 
To highlight the power of fitting multi-epoch data, the lower left panel shows the equivalent constraints on $M_\bullet$ and $a_\bullet$ (FoM $= 22.9$) obtained from a fit to a single epoch (row 13 in Table~\ref{tab:mock}) that has the same number of counts in its X-ray spectrum as do the previously considered sets of four spectra (rows 5-8 or rows 9-12) combined. The lower right panel shows the constraints on $M_\bullet$ and $a_\bullet$ (FoM $= 64.5$) obtained from a fit to four IMBH spectra (rows 14-17 in Table~\ref{tab:mock}). We set the exposure time of the four epochs to 50 ks. Multi-epoch spectra can partially break the degeneracy between $M_\bullet$ and $a_\bullet$;
for IMBHs, multi-epoch spectra can constrain both $M_\bullet$ and $a_\bullet$ quite well (cf.~the results of W21).}
\label{fig:multiepochs}
\end{figure*}

Figure \ref{fig:multiepochs} shows the constraints on $M_\bullet$ and $a_\bullet$ obtained when jointly fitting multi-epoch spectra together.  
For three different sets of spectra, we simultaneously fit all epochs with the model {\sc phabs*slimd}. We require the same $M_\bullet$, $a_\bullet$, and $\theta$ within each set of spectra, because those parameters are not expected to change between observations. The figure shows that a fit to multi-epoch spectra obtained when the mass accretion rate is nearly super-Eddington (upper left panel) yields tighter constraints\footnote{Note that the total number of counts in these four spectra is nearly the same; see Table \ref{fig:multiepochs}.} on both $M_\bullet$ and $a_\bullet$ than do multi-epoch spectral fits to only sub-Eddington epochs (upper right panel). The explanation for these tighter constraints is twofold. First, a higher accretion rate yields a harder spectrum, increasing the fraction of X-ray emission that falls in the observed band and thus also boosting its constraining power. Second, a higher accretion rate pushes the 
effective temperature maximum of the disk closer to the ISCO, where the influence of the BH spin has a stronger effect on the emergent X-ray emission. 

In our past work (W20; W21), fits to multi-epoch spectra constrained both $M_\bullet$ and $a_\bullet$ better than fits to individual, single-epoch spectra did. However, as multi-epoch spectra usually have more counts in total, 
we need to test whether the multi-epoch nature of multiple observations is intrinsically valuable for breaking parameter degeneracies.  To do this, we have generated a single mock spectrum with $\dot m=1$ and set the exposure time such that the spectrum has
the same number of counts as the four spectra combined in each of Figure \ref{fig:multiepochs}'s two upper panels. The lower panel of Figure \ref{fig:multiepochs} shows the $M_\bullet$ and $a_\bullet$ constraints from this single epoch spectrum. 
It is clear both by eye and by FoM analysis that this single-epoch, trans-Eddington spectrum provides weaker constraints than multi-epoch spectra including at least one trans-Eddington epoch (top left panel), but stronger constraints than multi-epoch spectra that only span sub-Eddington epochs.

The lower right panel shows the constraints on $M_\bullet$ and $a_\bullet$ from the fitting of the four epochs of simulated IMBH TDE spectra.
As we showed in Section \ref{sec:pd}, the model fits to a single epoch of a super-Eddington IMBH spectrum can already constrain both $M_\bullet$ and $a_\bullet$ reasonably well.  We generate four spectra with a 50 ks exposure time each (see rows 14 to 17 in Table~\ref{tab:mock}). The number of counts for each epoch is about $1/2$ to $1/3$ that in row 3 of Table~\ref{tab:mock} (our previous IMBH analysis, where the Eddington ratio was higher) and also in rows 5-12 (our previous multi-epoch analysis for SMBHs, which are usually brighter at fixed $\dot{m}$). 
We retrieve the input 
$M_\bullet$ and $a_\bullet$ values to within $1\sigma$ CL, and the CL contours are small relative to the other panels.
The constraints would be even better for slim disk fits to more highly super-Eddington accretion spectra, a situation which arises frequently for IMBH TDEs.

\subsection{Effect of edge-on disk}
\label{edg-on}
\begin{figure*}[ht!]
\gridline{\fig{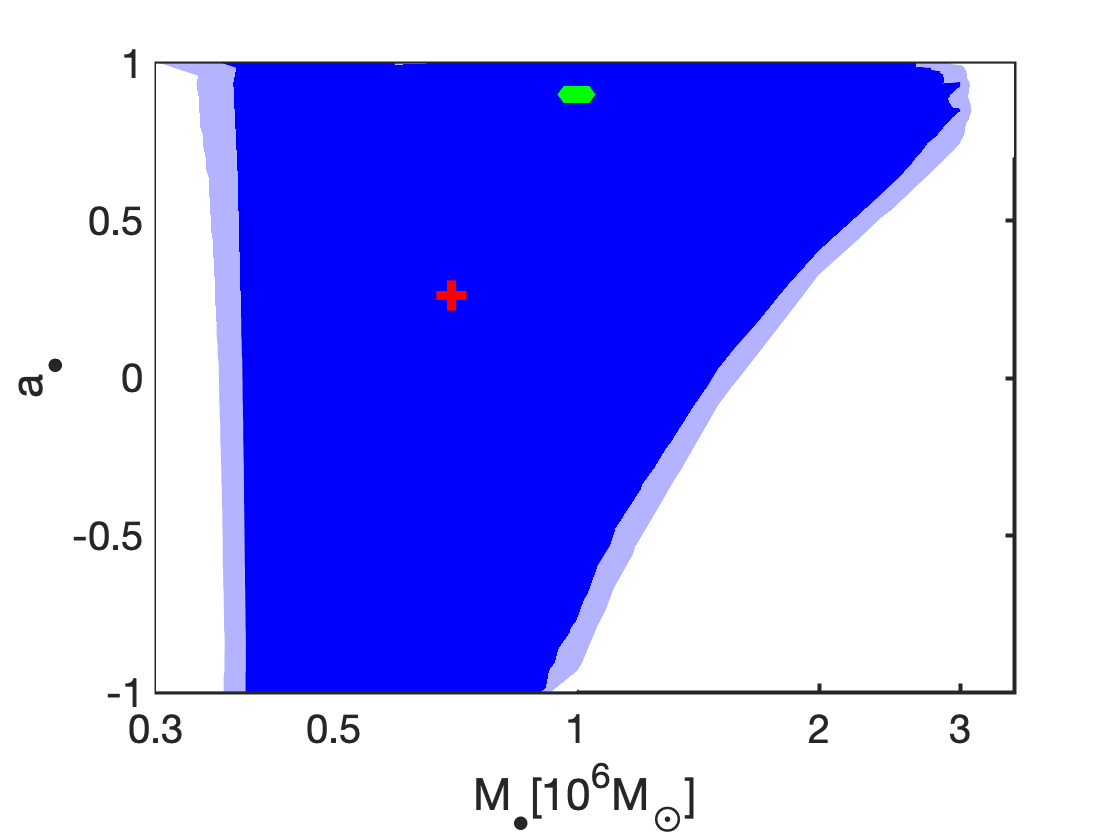}{0.45\textwidth}{$\dot m= 10$.}
\fig{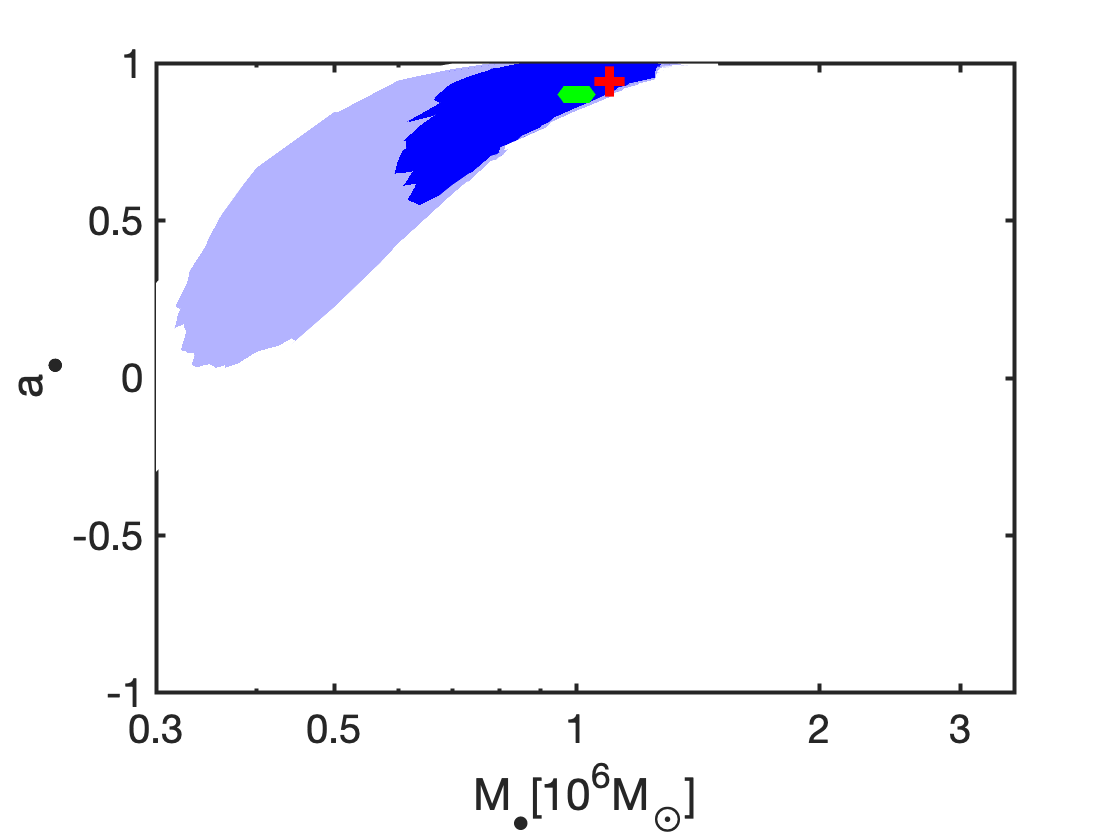}{0.44\textwidth}{$\dot m=$ 2, 1, 0.9, 0.8, and 0.7. }
}
\gridline{\fig{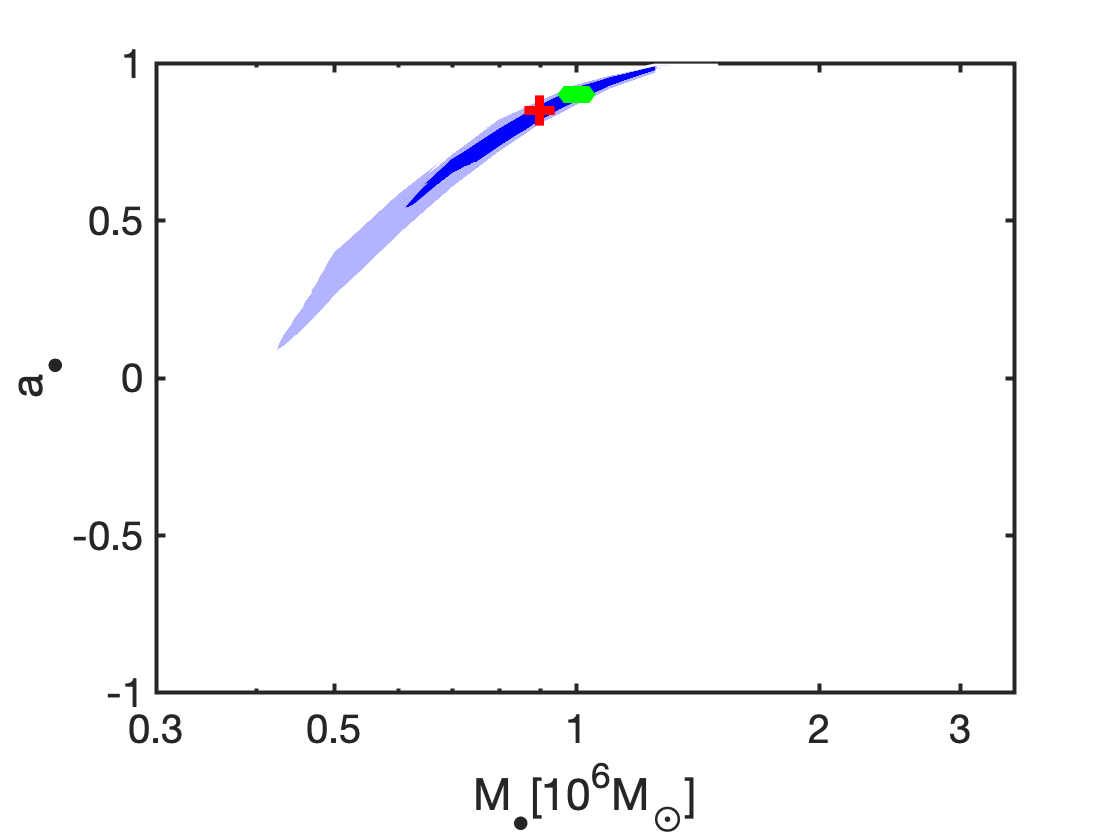}{0.45\textwidth}{$\dot m=$ 10, 2, 1, 0.9, 0.8, and 0.7.}
}
\caption{Constraints on $M_\bullet$ and $a_\bullet$ from fitting nearly edge-on ($\theta=80^\circ$) SMBH TDE spectra. The figure style is the same as in Fig. \ref{fig:multiepochs}.  The upper left panel shows the weak constraints on $M_\bullet$ and $a_\bullet$ from a high-$\dot{m}$ spectrum: $a_\bullet$ is 
unconstrained, while $M_\bullet$ is determined only to within a factor $\approx 10$. The upper right panel shows the improved constraints on $M_\bullet$ and $a_\bullet$ from fitting multi-epoch spectra with a wide range of mass accretion rates (and  $\theta=80^\circ$). Finally, the lower panel shows the constraints on $M_\bullet$ and $a_\bullet$ combining the 
six spectra used in both upper panels. The FoM of these three plots is 8.3, 94.6, and 233.7, respectively. The lower panel shows that an edge-on spectrum improves the joint constraint on $M_\bullet$ and $a_\bullet$ significantly, compared with the upper two panels.}
\label{fig:egdeon}
\end{figure*}

In this section, we explore how an edge-on 
inclination ($\theta$) of the inner disk with respect to our line of sight affects the constraints from model fits to multi-epoch spectra.
When the accretion rate decreases, the disk height is reduced and more X-rays from the inner disk have a free line of sight towards the observer. We generate six mock SMBH spectra with $\theta= 80^\circ$ and $\dot m$ ranging from 10 to 0.7. (See Table~\ref{tab:mock}, rows 18 to 23, for the values of other parameters.)
The 0.3-10 keV X-ray flux peaks for disks with $\dot m \approx 2$. For $\dot m >2$, the disk aspect ratio increases, and the X-ray emission of the inner disk is increasingly blocked by the disk edge.
For $\dot m <2$, the sky-integrated luminosity in the 0.3$-$10 keV band decreases quickly with decreasing mass accretion rate. 

Figure \ref{fig:egdeon} shows the constraints on $M_\bullet$ and $a_\bullet$ obtained by fitting our model spectral library to 
the mock observational spectra. The upper left panel shows the constraints on $M_\bullet$ and $a_\bullet$ from fitting a
single spectrum with $\dot m=10$. $M_\bullet$ is constrained to within a factor of $\approx$10 only, and there is no constraint on $a_\bullet$. The immediate reason for this is the very low number of X-ray counts (see Table \ref{tab:mock}), but the deeper explanation is that the X-ray photons from the inner disk are almost entirely blocked by the geometrically thick disk and are unable to reach the observer.  

The upper right panel shows the constraints on $M_\bullet$ and $a_\bullet$ obtained by fitting multi-epoch spectra. In this case, the multi-epoch fit to five spectra includes two with high accretion rates ($\dot m=2,1$), but not so high as to heavily obscure the inner, X-ray emitting disk annuli. As a result, both $M_\bullet$ and $a_\bullet$ are strongly constrained.

The lower panel shows the result combining all six spectra.
Interestingly, the addition of just one more spectrum improves the joint constraint on $M_\bullet$ and $a_\bullet$ by a factor of $\approx 2.5$ compared to the top right panel. This improvement arises because the additional, $\dot m=10$ spectrum provides a strong constraint on the disk inclination
similar to that shown in 
the lower right panel of Fig.\ref{fig:15oip}

\subsection{Effect of $N_{\rm H}$}
\begin{figure*}[ht!]
\gridline{\fig{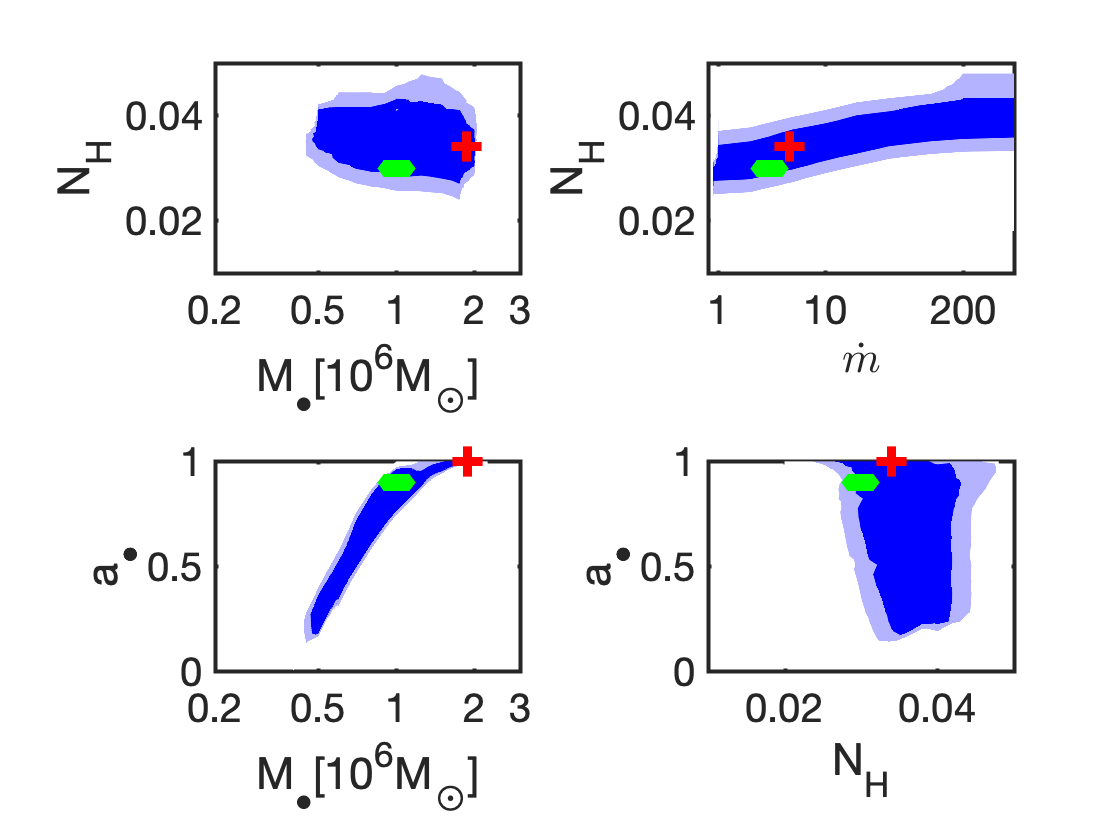}{0.5\textwidth}{$\dot m=$ 3.}
\fig{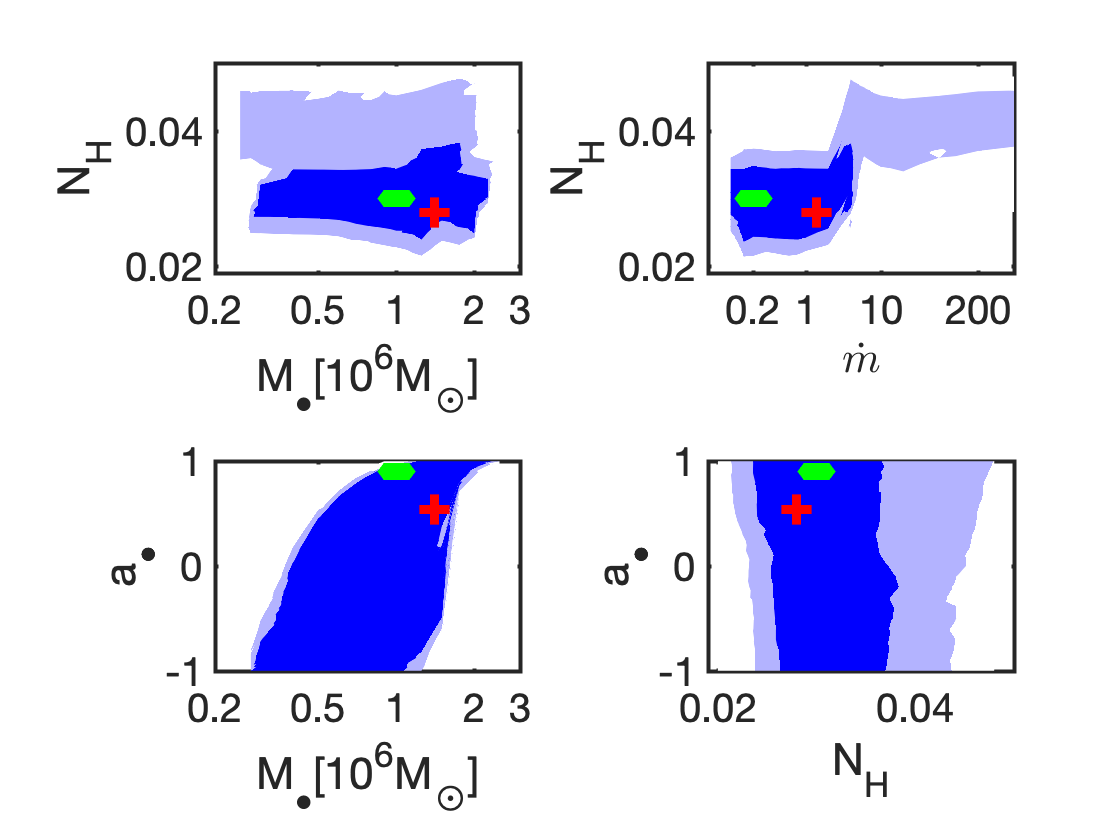}{0.5\textwidth}{$\dot m=$ 0.2.}
}
\caption{Effect of including the absorption $N_{\rm H}$ as a free parameter on the fitted results using a single spectral epoch, for two different SMBH accretion rates. The contours, as well as the 
colored points, are defined as in Fig. \ref{fig:multiepochs}.  The original disk parameters are $M_\bullet =10^6 M_\odot$, $a_\bullet=0.9$, $\theta =45^\circ$, and $N_{\rm H} = 0.03\times 10^{22} {\rm cm}^{-2}$. The left panels show the effect of $N_{\rm H}$ on a single super-Eddington spectrum ($\dot{m}=3$), while the right panels are for a single-epoch sub-Eddington spectrum ($\dot{m}=0.2$). These figures show that the input value of $N_{\rm H}$ can be recovered from the X-ray spectral fitting for both the super- and sub-Eddington cases examined here and that the uncertainty in $N_{\rm H}$ makes the constraints on other parameters weaker. }
\label{fig:NH1}
\end{figure*}

\begin{figure*}[ht!]
\gridline{\fig{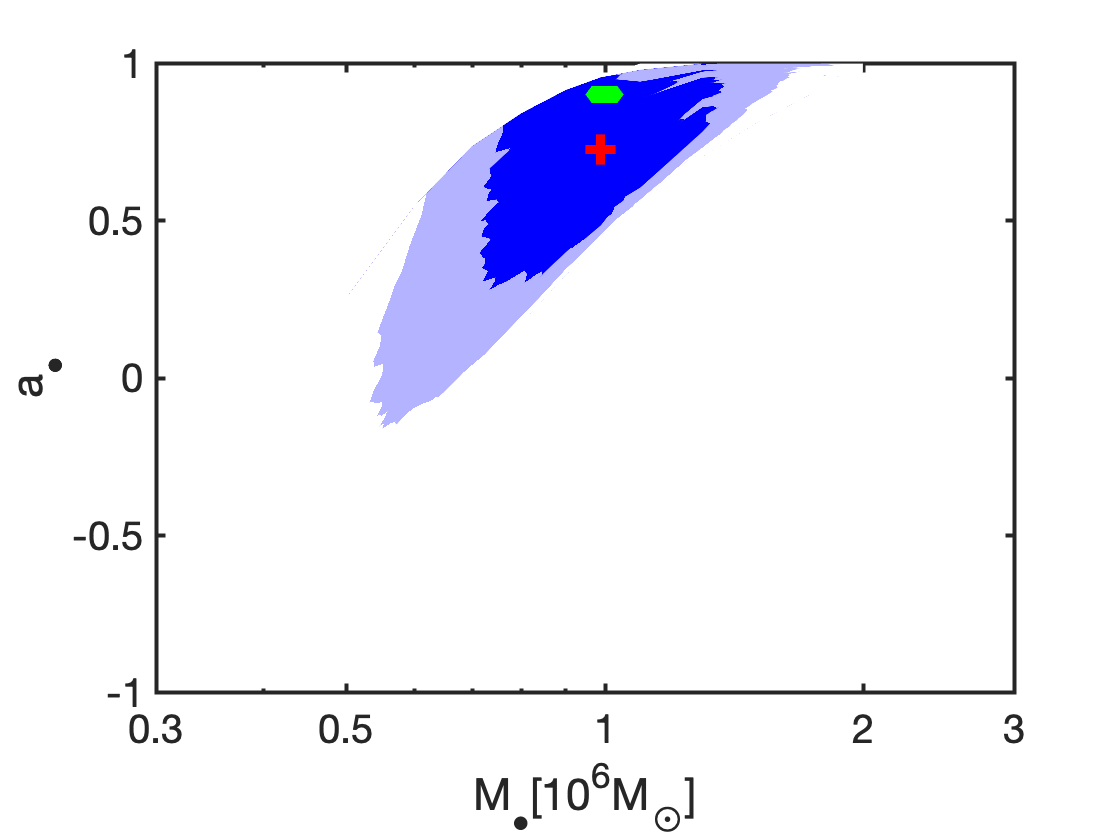}{0.49\textwidth}{$\dot m=$ 1, 0.9, 0.8, 0.7, 0.6, 0.5, 0.4, 0.3, and 0.2. Here $N_{\rm H,i}$ are fixed at $0.03\times 10^{22} {\rm cm}^{-2}$. }
\fig{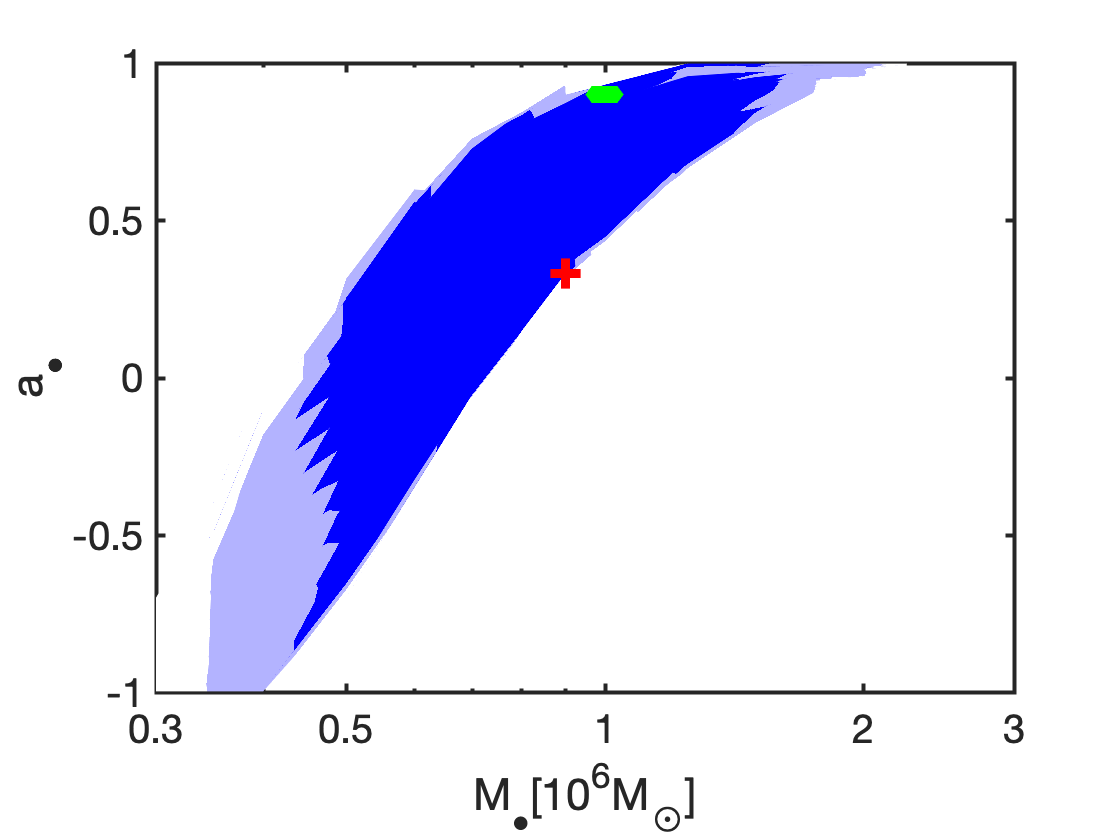}{0.5\textwidth}{$\dot m=$ 1, 0.9, 0.8, 0.7, 0.6, 0.5, 0.4, 0.3, and 0.2. Here $N_{\rm H,i}$ are free.}
}
\caption{Effect of including the absorption $N_{\rm H}$ as a free parameter on the fitted results from multiple-epoch observations. The figure style is the same as in Fig. \ref{fig:multiepochs}.  We fit mock spectra for the case of $M_\bullet = 10^6 M_\odot$, $a_\bullet = 0.9$, $\theta=45^\circ$, $N_{\rm H,i} = 0.03\times 10^{22} {\rm cm}^{-2}$, and various accretion rates.  The left panel shows the constraints on $M_\bullet$ and $a_\bullet$ for a fit to multi-epoch spectra where $N_{\rm H}$ is manually fixed at the true (input) value used for creating the mock spectra.  The right panel shows the equivalent constraints on $M_\bullet$ and $a_\bullet$ for the same spectra, except here $N_{\rm H}$ is not known {\it a priori}, and $N_{\rm H, i}$ are allowed to float freely in the fitting procedure for each epoch. These results show that the combined effect of the Milky Way, host galaxy, and intrinsic TDE absorption 
on the best-fit $M_\bullet$ and $a_\bullet$ constraints is significant. The FoM for the left and right panel is 76.3 and 27.0, respectively.}
\label{fig:NH2}
\end{figure*}

In this section, we quantify the importance of absorption on the X-ray spectrum. As the X-ray emission may be subject to absorption, the total amount of which is not known {\it a priori}, one needs to fit the absorption together with the other TDE parameters.  The absorbing column $N_{\rm H}$ may be portions of the interstellar medium located in the Milky Way or in the TDE host galaxy, in which case $N_{\rm H}$ should be the same from epoch to epoch.  Alternatively, a large fraction of $N_{\rm H}$ could arise from local gas produced in the TDE itself, in which case $N_{\rm H}$ would be expected to vary in time, likely experiencing an eventual decline.  This locally-produced gas could include bound but poorly circularized debris \citep{LoebUlmer97, Guillochon+14, Roth+16}, or 
outflowing\footnote{Note that even in situations where there is too little cool outflowing matter to produce a thermal optical/UV continuum \citep{MatsumotoPiran21}, frequency-dependent X-ray opacity may still strongly attenuate the observable disk emission \citep{MetzgerStone16}.} material \citep{MetzgerStone16, RothKasen18, Dai+18, LuBonnerot19, PiroLu20}.

Here we consider the effect of absorption on the best-fit parameters using fits both to a single-epoch spectrum and to multi-epoch spectra. As before, we 
define the mock spectrum with
an absorbing column of $N_{\rm H} = 0.03\times 10^{22} {\rm cm}^{-2}$, but we now allow
$N_{\rm H}$
to float in our fits, rather than specifying it as a prior.  Figure~\ref{fig:NH1} shows the effect of absorption on the parameter constraints from fits to single-epoch spectra (in one case super-Eddington, in the other sub-Eddington). From the contours of $\dot m=3$, we see that there is a degeneracy between 
$N_{\rm H}$ and 
$M_\bullet$, $a_\bullet$, and $\dot m$. As a result, we find poorer constraints on $M_\bullet$, $a_\bullet$ and $\dot m$ compared with the fit where $N_{\rm H}$ 
is known {\it a priori} (see plots with $\dot m=3$ in Figure~\ref{fig:ma}). However, the best-fit value of $N_{\rm H}$ is constrained to $0.035 \pm 0.01~( 10^{22}{\rm cm^{-2}})$, and thus the input value is regained at a 1$\sigma$ CL. 
From the CL contours on $\dot m=0.2$, we see that this degeneracy (of $N_{\rm H}$ with $M_\bullet$, $a_\bullet$, and $\dot m$) remains for sub-Eddington spectra, weakening constraints on parameters of interest. 
While the uncertainty in $N_{\rm H}$ can affect constraints on other disk parameters, $N_{\rm H}$ itself can be well-fit for both super- and sub-Eddington cases, at least for the values we have explored here.

Figure~\ref{fig:NH2} shows the effect of allowing the absorption parameter to float freely when fitting multi-epoch data. The left panel shows the constraints on $M_\bullet$ and $a_\bullet$ with fixed $N_{\rm H,i}$, while the right panel 
shows the
$M_\bullet$ and $a_\bullet$
constraints from fitting the same spectra but allowing $N_{\rm H,i}$ to float freely. We see that the additional parameters $N_{\rm H,i}$ make the constraints on $M_\bullet$ and $a_\bullet$ significantly worse at both the 1$\sigma$ and 2$\sigma$ level.  The substantial degeneracies of absorption with parameters of (generally) greater interest may motivate other approaches for constraining this ``nuisance parameter,'' such as fitting X-ray absorption lines \citep{Miller+15}, although this goes beyond the scope of the present work.

\section{Summary \& conclusions}
\label{conclusions}
X-ray observations of TDEs have a rich history.  They are not only the first means by which TDEs were detected (e.g., \citealt{Bade+96, KomossaBade99}), but by probing regions of spacetime close to the event horizon of the black hole, they have the potential to measure fundamental parameters such as black hole mass $M_\bullet$ and spin $a_\bullet$.  Our recent work (W20, W21) aimed to systematize the process of parameter estimation in TDE X-ray spectra, but one key problem was computational expense.  Applying full, on-the-fly general relativistic ray-tracing calculations to estimate confidence contours for the $\{M_\bullet, a_\bullet\}$ plane can take over $\sim 10^4$ CPU hours.  In this paper, we have addressed this problem by constructing a pre-tabulated library of slim disk model spectra that can be used by the community to fit soft X-ray observations of TDEs over only $\sim 10$ CPU hours. 

We use the model of W21 to construct this tabulated model. First, we solve the general relativistic stationary slim disk model, including the effect of angular momentum loss by radiation. Second, we use the relativistic ray-tracing code to calculate synthetic spectra. In these two steps, we have included gravitational redshift, Doppler, and lensing effects self-consistently. When calculating the spectrum, we determine the spectral hardening factor $f_c$ using Eq.~\ref{fc2}, and set the inner and outer disk edge to the ISCO and $2R_t$, respectively (except for very high-mass SMBHs). There are four free parameters for the model: $M_\bullet$, $a_\bullet$, $\dot m$, and the inclination $\theta$. Their ranges are listed in Table~\ref{tab:prior}. We employ linear interpolation to calculate the spectra with parameters between the grid points.
We evaluate the error on the calculated fluxes caused by the interpolation. 

We refit the X-ray spectra of ASASSN-14li and ASASSN-15oi with the library of model spectra and compare the results with those presented in W20. Using simulated mock X-ray spectra, we explore the full range of degeneracies between model parameters. We further evaluate how multi-epoch X-ray spectra, an edge-on disk configuration, and absorption can affect the constraints on parameters.  Our conclusions are:
\begin{enumerate}

\item
The error on the flux in the 0.2--10 keV band caused by linear interpolation is typically $<5\%$. The error increases for higher energy photons and can rise to $10\%$ far out on the Wien tail, when the flux is a factor of $10^6$ lower than at the peak of the X-ray spectrum. The error can be even larger, up to $40\%$, for an edge-on, high $\dot{m}$ disk configuration (see Figure \ref{fig:er1}). The constraints on $M_\bullet$ and $a_\bullet$ for ASASSN-14li and ASASSN-15oi obtained using the library of model spectra to fit the data are consistent with those from the full ray tracing model (W20). We also confirm the decrease found by W20 in $N_{\rm H}$ with time at the $>2\sigma$ CL for ASASSN-14li and at $>3\sigma$ for ASASSN-15oi.

\item
Both super- and sub-Eddington spectra can constrain $M_\bullet$. High spin, low BH mass, and/or highly super-Eddington spectra allow a stronger constraining power on $\{M_\bullet, a_\bullet\}$ than low spin, high BH mass, and/or sub-Eddington spectra (see Figure~\ref{fig:ma}). Typically, a single-epoch spectrum at a highly super-Eddington mass accretion rate can constrain $\{M_\bullet, a_\bullet\}$ well.

\item
The constraining power of $M_\bullet$ and $a_\bullet$ increases as a power law with the number of spectral counts. The index of this power law is higher for high $\dot{m}$, higher $a_\bullet$, and low $M_\bullet$.

\item
Multi-epoch spectra can help to break degeneracies in parameter estimation (see Figure~\ref{fig:multiepochs}). Nevertheless, multiple epochs of sub-Eddington mass accretion rate observations provide weaker constraints on $\{M_\bullet, a_\bullet\}$ than do super-Eddington epochs with similar total counts.

\item
Fitting a single spectrum 
of an edge-on inclination system constrains $\{M_\bullet, a_\bullet\}$ poorly, 
but simultaneous, multi-epoch spectral fitting
can place much tighter constraints 
(see Figure~\ref{fig:egdeon}).

\item
The level of absorption $N_{\rm H}$ can be recovered from fitting both super-Eddington and sub-Eddington mass accretion rate spectra. However, the uncertainty in $N_{\rm H}$
adversely affects the constraints obtained on $M_\bullet$ and $a_\bullet$. 

\end{enumerate}

Our relativistic slim disk library provides a systematic, computationally efficient framework for fitting and interpreting the growing wealth of observational data.
As noted in our previous papers (W20, W21), our model disks are inherently idealized, because complete, first-principles models for the hydrodynamic evolution of TDEs do not yet exist (see e.g., \citealt{BonnerotStone21} for a recent review). Most notably, at early times, TDE disks may possess substantial eccentricity or tilt that we currently ignore. Likewise, the slim disk solutions could break down for highly super-Eddington epochs.  

These potential problems are mitigated for sufficiently late-time observations, once the inner disk has fully circularized, internal and external torques have aligned it 
with the black hole equatorial plane, and the accretion rate has become sub-Eddington. These issues may also be reduced in certain regions of parameter space: higher spin values and more massive SMBHs likely favor fast alignment of initially tilted disks \citep{Franchini+16}, and deeply plunging TDEs may experience more rapid circularization \citep{Andalman+22}.  If rapid circularization is a pre-requisite for high X-ray luminosity, as has been suggested in past work \citep{Dai+15}, then residual disk eccentricity may not be a problem for most X-ray bright TDEs.  As the theory of TDEs progresses, the library framework we have constructed can be extended to incorporate more complex aspects of TDE disk physics and thereby eliminate some of these systematic uncertainties.  

The pre-tabulated model library presented here largely reduces the computational cost of fitting single- and multi-epoch TDE X-ray spectra with the slim disk model.  We anticipate that these publicly available tools will be more and more useful to the community as it transitions to an era of population studies on X-ray bright TDEs.  The {\it eROSITA} telescope has already found dozens of new X-ray TDEs \citep{Sazonov+21} and is likely to find hundreds if not thousands more before its mission is over \citep{Khabibullin+14, Jonker+20}.  The planned {\it Einstein Probe} mission has similar potential that may be realized this decade \citep{Yuan+15}.  This near-future wealth of X-ray TDEs will carry unprecedented information on the demographics of $\{M_\bullet, a_\bullet\}$ for massive black holes in quiescent galaxies, which can be explored efficiently with the framework we have introduced here.

\section*{Acknowledgements}

We thank Chi-Kwan Chan for his help in accessing computational resources and in constructing the model. We also thank Keith  Arnaud for his guidance for adjusting our model to the XSPEC table model. SW and AIZ thank Steward Observatory and the UA Department of Astronomy for post-doctoral support for SW. SW also thank RU Department of Astrophysics/IMAPP for post-docral support for SW. AIZ acknowledges additional support from NASA ADAP grant $\#80NSSC21K0988$.
NCS received support from the Israel Science Foundation (Individual Research Grant 2565/19). Our disk calculations were carried out at UA HPC systems, which are supported by the National Science Foundation under Grant No.~1228509. Our SED calculations were carried out at Open Science Grid \citep{Pordes+07,Sfiligoi+09}, which is supported by the National Science Foundation award $\#2030508$.  Discussions during the
2020 Yukawa Institute for Theoretical Physics (YITP) workshop
on ``Tidal Disruption Events: General Relativistic Transients'' at
Kyoto University were important to the completion of this work.

\begin{appendix}

\begin{figure}[ht!]
\plotone{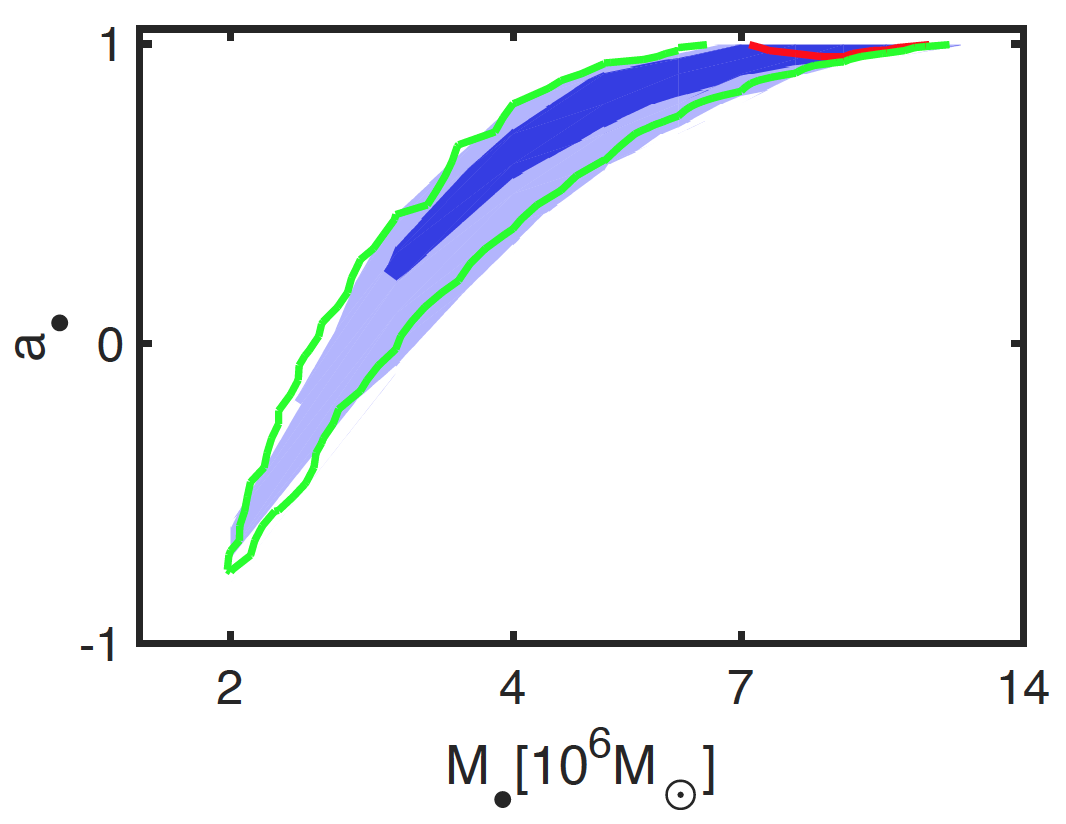}
\caption{Comparison of constraints on $M_\bullet$ and $a_\bullet$ for ASASSN-14li using the full ray tracing model from W21 and from W20.
The blue and light blue regions denote the $1\sigma$ and $2\sigma$ contours for fits from W20, while the red solid line and green solid line represent the $1\sigma$ and $2\sigma$ contours for fits from W21. Due to the angular momentum lost by radiation, the $1\sigma$ contour from W21 is much smaller than that from W20.}
\label{fig:14lima1}
\end{figure}
\section{Comparison of constraints on $M_\bullet$ and \lowercase{$a_\bullet$}
from W20 and W21}
\label{app:14lima}

In W21, we updated the slim disk solution by including the effect of angular momentum loss by radiation \citep{Abramowicz1996}. We found that this effect becomes strong for cases of high spin and low accretion rates, and can be ignored for low spin or high accretion rates. Here, we compare the constraints on $M_\bullet$ and $a_\bullet$ between the slim disk model of W21 and of W20, i.e., with and without the inclusion of angular momentum loss through radiation, respectively. We use the same spectra and parameter priors as in W20.

Figure \ref{fig:14lima1} shows the constraints on $M_\bullet$ and $a_\bullet$ with the slim disk model of W21 and of W20. Using the W21 model, we recover the best-fit and the 2$\sigma$ contour on the uncertainties for $M_\bullet$ and $a_\bullet$ from W20. However, the $1\sigma$ contour obtained from W21 is much smaller than for W20. This difference arises because the $\chi^2$ value of the best fit is smaller by about 1.0 for W21.
The decrease in $\chi^2$ can be understood as due to the high best-fit value of the BH spin making the inclusion of angular momentum loss through radiation important. 

\end{appendix}

\end{document}